\documentclass[conference]{IEEEtran}
\IEEEoverridecommandlockouts

\usepackage{cite}
\makeatletter
\renewcommand{\citepunct}{,\penalty\@m\hskip.13em plus .1em minus .1em\relax}

\makeatother

\usepackage{amsmath,amssymb,amsfonts}
\usepackage{textcomp}

\makeatletter
\def\@IEEEinterspaceratioM{0.29}
\def\@IEEEinterspaceMINratioM{0.21}
\def\@IEEEinterspaceMAXratioM{0.44}
\def\@IEEEinterspaceratioB{0.31}
\def\@IEEEinterspaceMINratioB{0.21}
\def\@IEEEinterspaceMAXratioB{0.48}
\makeatother

\usepackage[T1]{fontenc}
\usepackage[varqu]{zi4}

\usepackage{graphicx}
\usepackage[table]{xcolor}
\usepackage{xurl} 

\usepackage{booktabs}
\usepackage{makecell}

\usepackage{stfloats}

\usepackage{balance}

\usepackage{algorithm}
\usepackage{algorithmic}

\usepackage{enumitem}
\setlist[enumerate,1]{itemsep=1pt, topsep=1pt, leftmargin=12pt}
\setlist[itemize]{itemsep=1pt, topsep=1pt, leftmargin=8pt}

\usepackage{listings}
\usepackage{tikz}
\usetikzlibrary{arrows.meta, decorations.pathreplacing, fit}

\usepackage{comment}
\usepackage{xspace} 
\usepackage{microtype}
\usepackage[bookmarksnumbered,unicode,breaklinks]{hyperref}
\definecolor[named]{ACMPurple}{cmyk}{0.55,1,0,0.15}
\definecolor[named]{ACMDarkBlue}{cmyk}{1,0.58,0,0.21}
\hypersetup{
  colorlinks,
  linkcolor=ACMPurple,
  citecolor=ACMPurple,
  urlcolor=ACMDarkBlue,
  filecolor=ACMDarkBlue,
  pdflang={en},
  pdfdisplaydoctitle,
}

\usepackage{caption}
\usepackage{subcaption}
\renewcommand{\tablename}{Table}
\newcommand{\tm}{\texttrademark\xspace}
\newcommand{\myparagraph}[1]{\vspace{.1em}\noindent \textbf{#1}}
\newcommand{\cq}[1]{{\color{red} \{CQ: #1\}}}

\newcommand{\rebuttal}[1]{#1}
\newcommand{\Description}[2][]{}

\newcommand{\affmeta}{\IEEEauthorrefmark{1}}
\newcommand{\affamd}{\IEEEauthorrefmark{2}}

\makeatletter
\def\@IEEEauthorblockAtopspace{0.75ex}
\makeatother

\makeatletter
\long\def\@makefntext#1{\parindent 1em\noindent\hbox{\@makefnmark}#1}
\makeatother

\ifdefined\shownumbers  \fi

\begin{document}

\bstctlcite{IEEEtranBSTCTL:full-names}

\title{\huge Architecting for Hardware Diversity:
Meta's End-to-End Co-Design for High-Performance AI Training on AMD GPUs
}

\author{%
\IEEEauthorblockN{\begin{minipage}{\textwidth}\centering\small\linespread{1.05}\selectfont
Haoyu~Zhang\affmeta,
Arshad~Rahman\affamd,
Adnan~Aziz\affmeta,
Ajith~Prabhudev\affmeta,
Amin~Farmahini\affmeta,
Ben~Rhoads\affmeta,
Benson~Ma\affmeta,
Brian~Shi\affamd,
Cen~Zhao\affmeta,
CK~Luk\affmeta,
Dingming~Wu\affmeta,
Emad~El-Haraty\affmeta,
Eric~Chen\affmeta,
Frank~Li\affmeta,
Gary~Huang\affmeta,
Gefei~Zuo\affmeta,
Haoran~Zhang\affmeta,
Hiwot~Kassa\affmeta,
Hongyi~Zeng\affmeta,
Jack~Taylor\affamd,
Jeremy~Hadidjojo\affmeta,
John~Donati\affmeta,
Jonathan~Sheffield\affmeta,
Julio~Machado~Silva\affamd,
Karthik~Ganesan\affmeta,
Kent~Russell\affamd,
KR~Kishore\affmeta,
Lei~Tian\affmeta,
Liu~Ke\affmeta,
Muhammad~Osama\affamd,
Peng~Chen\affmeta,
Rakesh~Hemnani\affmeta,
Ryan~Swann\affamd,
Saleel~Kudchadker\affamd,
Shankar~Yanamandram\affmeta,
Shobhit~Kanaujia\affmeta,
Shucai~Xiao\affamd,
Srinath~Bayareddy\affmeta,
Srinivas~Vaidyanathan\affmeta,
Sudharssun~Subramanian\affmeta,
Supadchaya~Puangpontip\affmeta,
Taylor~Robie\affmeta,
Tianhao~Huang\affmeta,
Tyler~Hart\affmeta,
Viswesh~Sankaran\affmeta,
Weijun~Jiang\affamd,
Wenkai~Du\affamd,
Xiaodong~Wang\affmeta,
Xiaohu~Guo\affamd,
Yao~Rong\affmeta,
Yinbin~Ma\affmeta,
Yue~Dong\affmeta,
Yuzhen~Huang\affmeta,
Yvonne~Liu\affmeta,
Zhan~Shu\affmeta,
Dan~Rabinovitsj\affmeta,
Ajit~Mathews\affmeta,
Chunqiang~Tang\affmeta
\end{minipage}}
\IEEEauthorblockA{\small\textit{\affmeta Meta}\qquad\textit{\affamd AMD}}%
}

\maketitle
\ifdefined\shownumbers \thispagestyle{plain} \fi


\begin{abstract}

Meta deploys a heterogeneous fleet comprising NVIDIA GPUs, AMD GPUs, and proprietary MTIA accelerators to reduce costs and mitigate vendor lock-in. However, the extensive history of NVIDIA-specific optimizations hinders performance portability to other hardware. This paper presents our methods for optimizing alternative accelerators, using AMD Instinct\tm MI350X versus NVIDIA B200 for recommendation model training as a case study. We propose three reusable principles. First, hardware-model co-design is essential: rather than directly porting models from NVIDIA GPUs, revising model structures and software recipes allows full exploitation of accelerator-specific strengths (e.g., MI350X's $1.6\times$ HBM capacity relative to B200). Second, our analytical performance model enables zero-warmup optimal kernel selection across diverse datatypes and matrix shapes in production models, drastically simplifying tuning. Third, increasingly mature agentic workflows have lowered the barrier to porting from the CUDA ecosystem; we extend these workflows for microarchitecture-aware kernel optimizations and custom shim generation. Compared to the directly ported baseline, these methods deliver up to $1.5\times$ improvement in end-to-end training throughput. To our knowledge, this is the first hyperscaler report on productionizing AMD GPUs for large-scale training. Our contributions have been upstreamed to the open-source PyTorch, Triton, and ROCm ecosystems.

\end{abstract}

\begin{IEEEkeywords}
Accelerators, DLRM, GPU, kernels, machine learning, recommendation systems
\end{IEEEkeywords}

\section{Introduction}
\label{sec:intro}

The immense potential of AI applications has catalyzed an unprecedented expansion in the accelerator landscape. While NVIDIA remains the incumbent platform, the AMD Instinct\tm~\cite{amd-micro25,mi350x} GPU has emerged as a viable alternative. Meanwhile, hyperscalers have also developed custom silicon, including Meta MTIA~\cite{mtia-blog,mtia-gen1,athena,artemis}, Google TPU~\cite{tpu-lessons}, Amazon Trainium~\cite{aws-trainium}, Microsoft Maia~\cite{microsoft-maia}, Huawei Ascend~\cite{huawei-ascend}, Alibaba Hanguang~\cite{alibaba-hanguang}, and Baidu Kunlun~\cite{baidu-kunlun}.

Capitalizing on this hardware proliferation, Meta deploys a heterogeneous fleet comprising NVIDIA GPUs, AMD GPUs, and proprietary MTIA accelerators at scale. This strategic diversification delivers several benefits: (1) enabling cost optimization by preserving the flexibility to adopt the most economical solutions as the market evolves; (2) strengthening supply-chain resilience; and (3)~matching model characteristics to the most suitable hardware.

However, the extensive history of NVIDIA-specific optimizations hinders performance portability to other hardware. For example, fundamental differences in memory hierarchies and execution units (e.g., 32-thread warps vs. 64-thread wavefronts) thwart direct software migration between NVIDIA and AMD GPUs. These challenges are most acute in deep learning recommendation models (DLRMs)~\cite{arxiv19-dlrm,mlcommons2026-dlrmv3}. Unlike the relatively uniform large language models (LLMs), DLRMs are characterized by diverse model architectures, fragmented compute kernel sets, irregular matrix dimensions, and complex memory access patterns.

While industry adoption of AMD GPUs has largely focused on \emph{inference}—evidenced by Meta's deployment of hundreds of thousands of AMD GPUs—distributed \emph{training} presents a far greater technical challenge. This paper delivers the first hyperscaler report of enabling AMD GPUs for production-scale DLRM \emph{training}. We demonstrate that through systematic optimization, AMD MI350X achieves performance comparable to the incumbent NVIDIA B200 GPU across a diverse suite of production DLRMs. We distill our production experience into three reusable principles for new hardware integration:

\myparagraph{Hardware-model co-design.}
Different accelerator microarchitectures offer distinct advantages and trade-offs. Rather than directly porting models from NVIDIA GPUs, hardware-model co-design—revising the model structure and software recipe—is essential to leverage an accelerator's specific strengths. Specifically, we exploit MI350X's $1.6\times$ HBM capacity relative to B200. To maximize training efficiency, we expand per-GPU batch sizes and reduce activation checkpointing while simultaneously optimizing data or model parallelism and implementing efficient embedding re-sharding. Beyond training throughput, we utilize the additional HBM to enhance model quality by enabling longer input sequences, larger hidden dimensions, and expanded embedding tables. This holistic approach delivers a $20\%$ training speedup and a $>$0.1\% improvement in normalized entropy (NE)~\cite{workshop14-ne}, demonstrating that architectural synergies can simultaneously drive improvements in both hardware efficiency and model quality.

\myparagraph{Analytical performance model.}
Traditionally, selecting high-performance general matrix multiplication (GEMM) kernels relies on offline tuning—a cumbersome process that generalizes poorly across the disparate matrix dimensions common in DLRMs. We propose Origami, an analytical performance model that predicts optimal tile configurations in microseconds, achieving 93\% accuracy relative to exhaustively autotuned kernels across diverse matrix shapes. While analytical models are not new, Origami extends prior approaches to multi-chiplet architectures, with accuracy validated at production scale.

\myparagraph{Agentic workflows for kernel optimization.}
To reduce engineering burden and migration friction, we codify complex kernel-tuning principles into agentic workflows: automating microarchitecture-aware configuration selection, swizzling process IDs, incorporating occupancy hints, and leveraging specialized hardware instructions. Together, they achieve $1.6\times$--$2\times$ speedups across both hand-written and compiler-generated kernels. Furthermore, we leverage agentic workflows to automate the application of solutions that address hardware-specific bottlenecks; e.g., to mitigate the performance impact of atomics on MI350X, we apply a tree-reduction algorithm that utilizes shared memory to deliver up to $4\times$ speedup in custom kernels.

Our methods have been validated at scale, supporting production training on tens of thousands of MI350X GPUs at Meta. Across a representative set of production models, MI350X achieves a geometric-mean throughput of 101\% relative to our established B200 baseline---a platform highly optimized through extensive deployment. This performance parity validates the viability of our multi-vendor diversification strategy.

\myparagraph{Limitations.}
While end-to-end performance is comparable, MI350X's GEMM performance remains an area for improvement, due to both the overhead of emulated TF32 GEMMs in the absence of native hardware support and the relative maturity of NVIDIA cuBLAS over AMD hipBLAS in handling irregular matrix shapes.

Although we use DLRMs to drive our optimizations, the underlying principles are generalizable to a broader spectrum of ML workloads, including LLMs. Indeed, we have successfully optimized LLM training on clusters of more than 2,000 MI350X GPUs; a comprehensive analysis of LLM-specific performance is reserved for future work.

To our knowledge, this is the first hyperscaler report on production-scale training using AMD GPUs. Our purpose is not to rank accelerators, but to share generalizable optimization methods for emerging hardware by synthesizing insights from our large-scale deployment of AMD, NVIDIA, and MTIA platforms. Our contributions have been upstreamed to the open-source PyTorch, Triton, and ROCm ecosystems.

\section{Background}
\label{sec:background}

\subsection{DLRM Evolution}
\label{sec:bg-dlrm}

\begin{figure}
    \centerline{\includegraphics[width=.9\columnwidth]{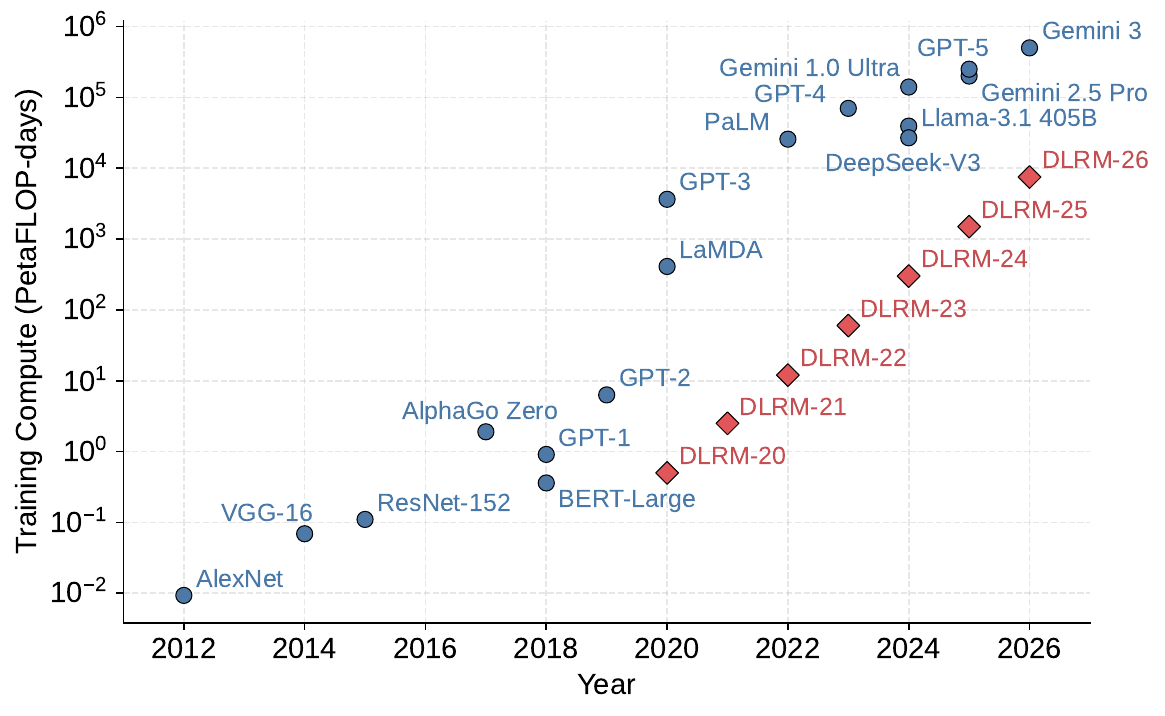}}
    \Description[Comparing deep learning models]{Total training compute (PetaFLOP-days) over time for major deep learning model generations, including LLMs and DLRMs.}
    \caption{Total training compute for deep learning models.}
    \label{fig:train-compute}
\end{figure}


Deep learning recommendation models (DLRMs)~\cite{arxiv19-dlrm} have become the canonical architecture for industrial recommendation and personalization systems across major technology companies, such as
Meta~\cite{isca22-dlrm},
ByteDance~\cite{recsys22-monolith},
Google~\cite{kdd16-widedeep}, and
Alibaba~\cite{kdd18-din}.
DLRMs often comprise (1) a \emph{sparse component} that uses embedding tables to map high-cardinality categorical signals---such as user IDs, item IDs, geographic regions, and language preferences---into dense vectors, and (2) a \emph{dense component} that learns feature interactions to predict user engagement (e.g., click-through or conversion rates). The importance of this model family is reflected both externally---through the MLPerf recommendation benchmark evolution and the adoption of DLRM~\cite{hpca20-dlrm}, DLRM-v2~\cite{isca22-dlrm}, and DLRM-v3~\cite{icml24-hstu,mlcommons2026-dlrmv3}---and internally through the scale of Meta's production deployment and infrastructure investment dedicated to improving DLRM efficiency.

\noindent \textbf{Sparse evolution.} Embedding tables grow rapidly in both vocabulary size and dimensionality, forcing distributed sharding and collective-heavy embedding exchange.
As training clusters expanded, two practical directions emerged. First, multi-stage pipelining became essential~\cite{recsys22-torchrec}: embedding lookup, all-to-all exchange, and optimizer steps must be organized to reduce exposed latency and avoid synchronization bubbles. Second, sparse parallelism itself evolved beyond single dimension. Recent work on two-dimensional sparse parallelism~\cite{arxiv25-2d-sparse} introduces a hierarchical strategy that adds data-parallel structure on top of model-parallel embedding sharding, reducing all-to-all scope, lowering embedding activation memory growth, and mitigating imbalance/straggler behavior. Together, these trends reflect a broader shift: the sparse subsystem is no longer ``just embeddings,'' but a distributed system whose parallelism strategy, pipeline structure, and optimizer semantics jointly determine scalability.


\noindent \textbf{Dense evolution.} Classical DLRMs pair large embedding tables with relatively shallow dense components (e.g., a few fully connected layers for feature interaction). Recent evolution of DLRMs (Fig.~\ref{fig:train-compute}) shows that scaling the dense subsystem can yield predictable quality gains analogous to LLM scaling laws~\cite{arxiv20-scalinglaws,arxiv22-chinchilla}. Wukong~\cite{icml24-wukong} demonstrates this by stacking factorization-machine layers with a synergistic upscaling strategy, achieving smooth quality improvements over orders of magnitude in dense-side FLOPs. A complementary trend is the adoption of Transformer-like modules to model long user-interaction histories. Meta's HSTU~\cite{icml24-hstu} exemplifies this sequence-centric direction, replacing shallow pooling with attention over hundreds of past events. Together, these advances open a second axis for improving recommendation quality: in addition to scaling embedding tables, practitioners can now scale dense computation (FLOPs) to capture richer feature interactions and temporal patterns.

\subsection{AMD MI350X vs NVIDIA B200}
\label{sec:bg-amd}

\begin{table*}[tbp!]
\centering
\footnotesize
\begin{tabular*}{\textwidth}{@{\extracolsep{\fill}}lcccc@{}}
\toprule
\textbf{Specification} & \textbf{MI350X} & \textbf{MI355X} & \textbf{B200} \\
\midrule
Architecture & CDNA 4 & CDNA 4 & Blackwell \\
Compute Units / Streaming Multiprocessors & 256 (8 XCD $\times$ 32) & 256 (8 XCD $\times$ 32) & 148 (2 chiplets) \\
Peak FP16/BF16 Dense Performance & 2300 TFLOPS & 2500 TFLOPS & 2250 TFLOPS \\
HBM Type / Capacity & HBM3e / 288 GB & HBM3e / 288 GB & HBM3e / 180 GB \\
HBM Bandwidth & Up to 8 TB/s & Up to 8 TB/s & Up to 8 TB/s \\
LDS per CU or SM & 160 KB & 160 KB & 256 KB \\
Scale Up BW per GPU (uni-dir) & 538 GB/s & 538 GB/s & 900 GB/s \\
Cooling & Air-cooled & Direct Liquid Cooled & Air-cooled \\
Deployed Power Limit & 1000 W & 1400 W & 750 W \\
Deployment & Meta On-Prem & Public Cloud & Meta On-Prem \\
\bottomrule
\end{tabular*}
\caption{Comparison of AMD Instinct MI350X, MI355X, and NVIDIA Blackwell B200 GPUs. MI355X is a 1400\,W design using direct liquid cooling, while Meta uses the 1000\,W air-cooled MI350X. 
Meta's deployment of B200 operates at 750\,W, which is different from the general B200 HGX solution that operates at 1000\,W. \rebuttal{Meta operates both GPUs below their vendor-specified TDPs on the same air-cooled platform. B200 is capped at a lower power limit to enable air cooling because its power density is higher than that of MI350X.}
}
\label{tab:amd-gpu-comparison}
\end{table*}


Table~\ref{tab:amd-gpu-comparison} compares the GPU specs. AMD MI350X is built on the CDNA\tm 4 architecture and is the first AMD GPU used for production-scale training at Meta. MI350X features eight accelerator complex dies (XCDs) with 32 compute units (CUs) each (a total of 256 CUs), a 288\,GB HBM3e memory pool with up to 8\,TB/s bandwidth, and 538\,GB/s of scale-up bandwidth per GPU. B200 offers 180\,GB of HBM capacity, with 900\,GB/s of scale-up bandwidth per GPU.

\rebuttal{B200 and MI350X are same-generation flagships with closely matching theoretical bounds: their near-identical peak compute ($2250$ vs.\ $2300$\,BF16 TFLOPS) and HBM bandwidth ($8$\,TB/s) yield largely coincident roofline curves~\cite{cacm09-roofline}, and thus a similar performance envelope for ML workloads. Their strengths are complementary: B200 leads on per-GPU scale-up bandwidth ($900$ vs.\ $538$\,GB/s), an all-to-all NVSwitch fabric (versus AMD's point-to-point xGMI mesh), shared memory per SM ($256$ vs.\ $160$\,KB), and native TF32 support, whereas MI350X's principal advantages are its $1.6\times$ HBM capacity ($288$ vs.\ $180$\,GB) and higher deployed power.} 

\begin{figure}
    \centerline{\includegraphics[width=\columnwidth]{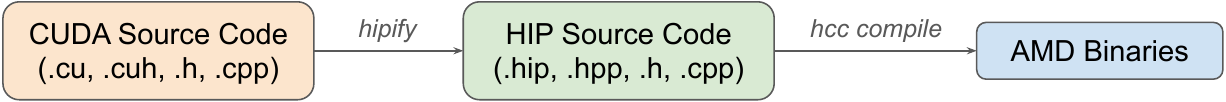}}
    \caption{\textit{Hipification} process to enable ML jobs on AMD GPUs.}
    \label{fig:hipification}
\end{figure}

\myparagraph{Hipification.} \textit{Hipification} is AMD's CUDA-to-HIP (Heterogeneous-compute Interface for Portability) translation process that converts NVIDIA CUDA code into HIP equivalents; see Fig.~\ref{fig:hipification}. There are two primary approaches: (1) compile-time hipification using AMD's open-source \texttt{hipify-clang} compiler, which translates CUDA constructs to HIP at build time, and (2) rule-based source-code translation using \texttt{hipify\_torch}, which traverses and rewrites source files before compilation. While hipification provides broad functional compatibility and enables most CUDA workloads to run on AMD with minimal manual effort, performance portability remains a challenge. For instance, CUDA uses a warp size of 32, whereas an AMD GPU uses a wavefront size of 64. Kernels that naively hardcode warp sizes will likely perform suboptimally without tuning.

The emergence of agentic AI tools has fundamentally lowered the cost of switching from NVIDIA to AMD. At Meta, we leverage AI agents to address issues in the hipification process. These agents iteratively compile code, interpret error messages, and apply fixes until the build succeeds. When a niche CUDA library lacks a ROCm equivalent, AI agents can even generate custom shims to mimic the missing functionality. This makes AMD GPUs accessible to a broad range of ML researchers, who can now deploy high-performance code without mastering low-level hardware nuances.



\myparagraph{Meta's Usage of AMD GPUs.}
Meta started using the AMD MI300X~\cite{meta-oci} for inference in 2024. Today, models compatible with AMD GPUs make up over 60\% of production model types, weighted by resource footprint.
Training DLRMs, on the other hand, presents a greater challenge than inference:



\begin{itemize}[nosep]
    \item \textbf{Synchronization and network pressure:} Training involves numerous GPUs in tight lockstep with frequent synchronization, heavily straining network efficiency.
    \item \textbf{Kernel and optimizer:} Training requires comprehensive backward kernels and advanced optimizer support.
    \item \textbf{Generalization:} While inference can be optimized per workload, training must rely on generalized optimizations for out-of-the-box performance—orders of magnitude more training trials run before a model is finalized, making per-job optimization infeasible.
\end{itemize}

\noindent Addressing these challenges requires systematic optimizations across the software stack, from low-level libraries (Section~\ref{sec:rocm}) and kernel tuning (Section~\ref{sec:kernel}) to high-level model co-design (Section~\ref{sec:model}).

\section{Roofline-Driven Optimization of ROCm Libraries}
\label{sec:rocm}

The ROCm software platform provides optimized libraries, runtimes, and firmware for AMD GPUs. Grounded in roofline models~\cite{cacm09-roofline} and hardware constraints, we optimize core GEMM kernels and selection algorithms to approach theoretical limits. 
To optimize RCCL, the collective communication library for distributed training, we derive theoretical bounds from hardware specifications as optimization targets. These roofline-driven principles enable low-level libraries to deliver robust, generalizable performance across diverse AI workloads.

\subsection{GEMM Kernels for AMD MI350X}

DLRM workloads present a broad range of matrix shapes and sizes, including irregular dimensions that stress standard GEMM kernel selection strategies. To deliver strong performance across this diversity, we integrate performance enhancements into hipBLASLt~\cite{AMD:hipBLASLt} GEMM kernels through its assembly backend, TensileLite~\cite{AMD:Tensile:2018}. These enhancements leverage new CDNA 4~\cite{AMD:CDNA4:2025} hardware capabilities alongside algorithmic refinements that reduce the solution space and enable finer-grained runtime parameter selection via Origami (Section~\ref{sec:origami}). Beyond exploiting new ISA features, we employ the following algorithmic techniques.

\myparagraph{Stream-K} partitions GEMM computations across CUs at finer granularity than traditional data-parallel tiling, improving load balance for tensor sizes that do not evenly tile the device. We also introduce new reduction strategies and runtime dispatch logic, improving robustness across a broad range of compute- and memory-bound tensor sizes. Notably, a single GEMM kernel can now achieve performance previously requiring multiple specialized kernels.

\myparagraph{Workgroup mapping via space-filling curves.}
MI350X GPUs use a multi-XCD architecture in which workgroups are distributed across XCDs, each with a private L2 cache. Scheduling workgroups operating on adjacent or overlapping tiles across different XCDs degrades cache locality and efficiency. To address this, hipBLASLt introduces a mapping strategy based on space-filling curves that preferentially co-locates workgroups processing nearby tiles on the same XCD, improving spatial locality, cache utilization, and power efficiency. Multiple mapping strategies are supported within the GEMM assembly kernels, with runtime-selectable parameters integrated with Origami to select the optimal strategy for a given tensor size, kernel configuration, and target hardware.

\subsection{Origami: Analytical Kernel Selection}
\label{sec:origami}

Prior kernel selection approaches in hipBLASLt relied on offline autotuning: enumerating kernel configurations on representative tensor shapes and storing the best selections in lookup tables. This approach fails at scale: recommendation models execute thousands of unique GEMM shapes per model, and maintaining exhaustive databases for every shape is impractical.

Origami replaces offline autotuning with an analytical performance model that predicts optimal tile configurations in microseconds. The model can be viewed as inferring the critical path of a GEMM problem across the memory and compute hierarchy. It captures the full tiling hierarchy---from matrix instruction dimensions through local data share (LDS) tiles to implicit cache tiles formed by workgroups sharing the L2 and last level cache (LLC)---and computes cycle counts geometrically at each level.

\myparagraph{Performance Model.}
We consider a GEMM operation $C = A \times B$, where $A \in \mathbb{R}^{M \times K}$, $B \in \mathbb{R}^{K \times N}$, and $C \in \mathbb{R}^{M \times N}$. To efficiently execute this operation on the GPU, the output matrix $C$ is partitioned into tiles, each computed by a workgroup. A workgroup iterates along the $K$ dimension, loading tiles of $A$ and $B$, performing multiply-accumulate operations, and writing the result back to global memory. The goal of the performance model is to select the tile configuration $(M_T, N_T, K_T)$ that minimizes total execution latency.

For each candidate tile configuration with LDS tile size $(M_T, N_T, K_T)$, Origami computes cycle counts for compute and memory operations, then selects the configuration minimizing total latency. The compute latency for one tile is:
\begin{equation}
L_\text{compute} = N_\text{MI} \cdot L_\text{MI}
\end{equation}
where $N_\text{MI} = \lceil M_T/M_I \rceil \cdot \lceil N_T/N_I \rceil \cdot \lceil K_T/K_I \rceil$ is the number of matrix instructions per tile (macro tile dimensions divided by instruction dimensions) and $L_\text{MI}$ is the architecture-specific instruction latency.

Memory latency is computed by cascading traffic through the cache hierarchy. The bytes loaded per tile are:
\begin{equation}
\text{Bytes}_\text{tile} = M_T \cdot K_T \cdot b_A + N_T \cdot K_T \cdot b_B
\end{equation}
where $b_A, b_B$ are bytes per element. Origami predicts cache hit rates by modeling workgroup sharing within each memory scope. For $N_\text{CU}$ CUs sharing a cache, arranged as an $M_C \times N_C$ grid of output tiles, the predicted hit rate is:
\begin{equation} \label{eq:hit-rate}
H = 1 - \frac{M_C \cdot M_T + N_C \cdot N_T}{M_C \cdot N_C \cdot (M_T + N_T)}
\end{equation}
This formula applies to L2 and LLC, with $N_\text{CU}$ set to the number of CUs sharing that cache scope. It captures data reuse: $A$ tiles are shared across $N_C$ workgroups computing different output columns, while $B$ tiles are shared across $M_C$ workgroups computing different rows. Tiles are rounded to the nearest cache line width in the contiguous dimension of the allocation (determined by transpose). The model constrains cache tile dimensions when working sets exceed cache capacity.

Traffic reaching each memory level determines hierarchy latency. Bytes cascade through miss rates: $\text{Bytes}_\text{LLC} = (1-H_\text{L2}) \cdot \text{Bytes}_\text{total}$ and $\text{Bytes}_\text{HBM} = (1-H_\text{LLC}) \cdot \text{Bytes}_\text{LLC}$, where each $H$ is computed using the CU count for that scope. Memory latency in cycles is:
\begin{equation}
L_\text{mem} = \max\left(\frac{\text{Bytes}_\text{L2}}{R_\text{L2}},\; \frac{\text{Bytes}_\text{LLC}}{R_\text{LLC}},\; \frac{\text{Bytes}_\text{HBM}}{R_\text{HBM}}\right)
\end{equation}
where $R$ is bytes per compute cycle at each level. \rebuttal{Expressing memory bandwidth in bytes per compute cycle keeps the memory and compute latencies in the same unit, so the model need not reconcile the GPU's separate memory and compute clock frequencies.}

\myparagraph{Issue Rate Bound.}
Effective memory throughput can be limited by the rate at which CUs issue loads and stores, not just bandwidth. This creates another parallelism-locality tradeoff on what are usually considered bandwidth-bound problems: smaller tiles increase the number of vector lanes issuing load/store requests but reduce data reuse, while larger tiles improve locality at the cost of issue rate. Origami models achievable throughput as a function of occupied vector lanes to navigate this tradeoff.

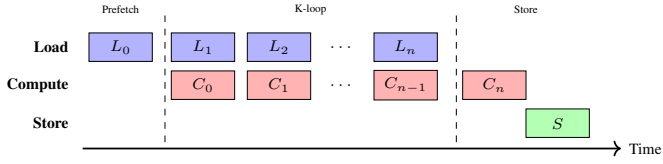
\begin{figure}[t]
\centering
\resizebox{\columnwidth}{!}{%
\begin{tikzpicture}[
    box/.style={draw, minimum width=1.0cm, minimum height=0.45cm, font=\scriptsize},
    load/.style={box, fill=blue!30},
    compute/.style={box, fill=red!30},
    store/.style={box, fill=green!30},
    label/.style={font=\scriptsize\bfseries}
]
\draw[->, thick] (0,-.4) -- (8.5,-.4) node[right] {\scriptsize Time};

\node[label, anchor=east] at (-0.1, 1.2) {Load};
\node[label, anchor=east] at (-0.1, 0.6) {Compute};
\node[label, anchor=east] at (-0.1, 0) {Store};

\node[load] at (0.6, 1.2) {$L_0$};
\draw[dashed] (1.3, -0.3) -- (1.3, 1.7);
\node[font=\tiny] at (0.6, 1.8) {Prefetch};

\node[load] at (1.9, 1.2) {$L_1$};
\node[compute] at (1.9, 0.6) {$C_0$};

\node[load] at (3.1, 1.2) {$L_2$};
\node[compute] at (3.1, 0.6) {$C_1$};

\node at (4.1, 1.2) {\scriptsize $\cdots$};
\node at (4.1, 0.6) {\scriptsize $\cdots$};

\node[load] at (5.1, 1.2) {$L_{n}$};
\node[compute] at (5.1, 0.6) {$C_{n-1}$};

\draw[dashed] (5.9, -0.3) -- (5.9, 1.7);
\node[font=\tiny] at (3.6, 1.8) {K-loop};

\node[compute] at (6.5, 0.6) {$C_n$};
\node[store] at (7.5, 0) {$S$};
\node[font=\tiny] at (7.0, 1.8) {Store};

\end{tikzpicture}%
}
\caption{Pipelined GEMM execution. Loads/stores use vector units; compute uses matrix units. The K-loop overlaps load and compute.}
\label{fig:gemm-pipeline}
\end{figure}

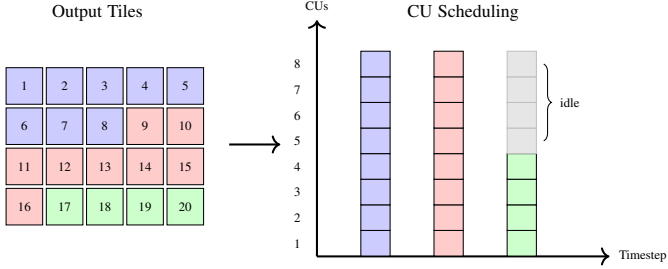
\begin{figure}[t]
\centering
\resizebox{\columnwidth}{!}{%
\begin{tikzpicture}[
    tile/.style={draw, minimum width=0.5cm, minimum height=0.5cm, font=\tiny},
    t1/.style={tile, fill=blue!20},
    t2/.style={tile, fill=red!20},
    t3/.style={tile, fill=green!20},
    idle/.style={tile, fill=gray!20, draw=gray!50},
    label/.style={font=\scriptsize}
]

\node[label] at (1.5, 2.8) {Output Tiles};
\foreach \i in {0,...,3} {
    \foreach \j in {0,...,4} {
        \pgfmathtruncatemacro{\idx}{\i*5 + \j + 1}
        \ifnum\idx<9
            \node[t1] at (.5+\j*0.55, 1.8-\i*0.55) {\idx};
        \else
            \ifnum\idx<17
                \node[t2] at (.5+\j*0.55, 1.8-\i*0.55) {\idx};
            \else
                \node[t3] at (.5+\j*0.55, 1.8-\i*0.55) {\idx};
            \fi
        \fi
    }
}

\draw[->, thick] (3.3, 1.0) -- (4.0, 1.0);

\node[label] at (6.5, 2.8) {CU Scheduling};

\draw[->, thick] (4.5, -0.53) -- (8.5, -0.53) node[right] {\tiny Timestep};
\draw[->, thick] (4.5, -0.53) -- (4.5, 2.7) node[above] {\tiny CUs};

\foreach \i in {1,...,8} {
    \node[font=\tiny, anchor=east] at (4.4, \i*0.35 - 0.7) {\i};
}

\foreach \i in {1,...,8} {
    \node[t1, minimum width=0.4cm, minimum height=0.35cm] at (5.3, \i*0.35 - 0.7) {};
}
\node[font=\tiny] at (5.3, -0.8) {$\omega_1$};

\foreach \i in {1,...,8} {
    \node[t2, minimum width=0.4cm, minimum height=0.35cm] at (6.3, \i*0.35 - 0.7) {};
}
\node[font=\tiny] at (6.3, -0.8) {$\omega_2$};

\foreach \i in {1,...,4} {
    \node[t3, minimum width=0.4cm, minimum height=0.35cm] at (7.3, \i*0.35 - 0.7) {};
}
\foreach \i in {5,...,8} {
    \node[idle, minimum width=0.4cm, minimum height=0.35cm] at (7.3, \i*0.35 - 0.7) {};
}
\node[font=\tiny] at (7.3, -0.8) {$\omega_3$};

\draw[decorate, decoration={brace, amplitude=3pt, mirror}] (7.6, 1.05) -- (7.6, 2.1);
\node[font=\tiny, anchor=west] at (7.7, 1.575) {idle};

\end{tikzpicture}%
}
\caption{Wave quantization tail effect. A 4$\times$5 output (20 tiles) on 8 CUs requires 3 timesteps with $20 \bmod 8 = 4$ underutilized CUs in the final timestep. Origami addresses this via two strategies: tile selection to change the number of output tiles, and Stream-K to partition work along the K dimension.}
\label{fig:wave_quant}
\end{figure}

\myparagraph{Latency Model.}
Fig.~\ref{fig:gemm-pipeline} illustrates the pipeline structure: the prologue issues the initial load, the steady-state K-loop overlaps loads with compute, and the epilogue completes the final compute and store. Per-tile latency combines these phases:
\begin{equation}
L_\text{tile} = L_\text{prefetch} + \bigl\lceil K/K_T \bigr\rceil \cdot \max(L_\text{compute}, L_\text{mem}) + L_\text{store}
\end{equation}
The number of output tiles $T_\text{out} = \lceil M/M_T \rceil \cdot \lceil N/N_T \rceil$ determines timestep count $\omega = \lceil T_\text{out} / N_\text{CU} \rceil$. Total latency is $L_\text{total} = \omega \cdot L_\text{tile}$. Fig.~\ref{fig:wave_quant} illustrates the tail effect: when tiles do not evenly divide across CUs, the final timestep underutilizes the GPU.

This static analysis is enabled by the regularity of GEMM workloads: uniform tile shapes and predictable access patterns. Origami evaluates all candidate configurations and selects the one minimizing $L_\text{total}$. Each candidate configuration evaluates in nanoseconds; the full search across all candidates completes in microseconds.

\myparagraph{Grid Size Selection.}
Beyond tile configuration, Origami selects the optimal Stream-K grid size. Stream-K partitions work in the K dimension across CUs, and Origami balances three considerations: parallelism to maximize CU utilization, locality to enable data reuse across CUs iterating over distinct K slices, and scheduling to load balance across XCDs and minimize the wave quantization problem at each level of the chiplet hierarchy.

\myparagraph{Chiplet-Aware Mapping.}
MI350X comprises 8 XCDs, each with private L2 cache. Origami extends hierarchical tiling to the chiplet level: a device-scope tile is composed of 8 XCD-scope tiles, each containing workgroups that share L2. XCD-aware workgroup ID swizzling maps spatially adjacent output tiles to the same XCD, improving L2 hit rates. Origami also selects the L2 tile shape dynamically based on the number of CUs per chiplet, arranging the concurrent workgroups into an $M_C \times N_C$ grid to maximize the cache hit rate predicted by Equation~\ref{eq:hit-rate}.

\begin{figure}[t]
    \centering
    \includegraphics[width=.8\columnwidth]{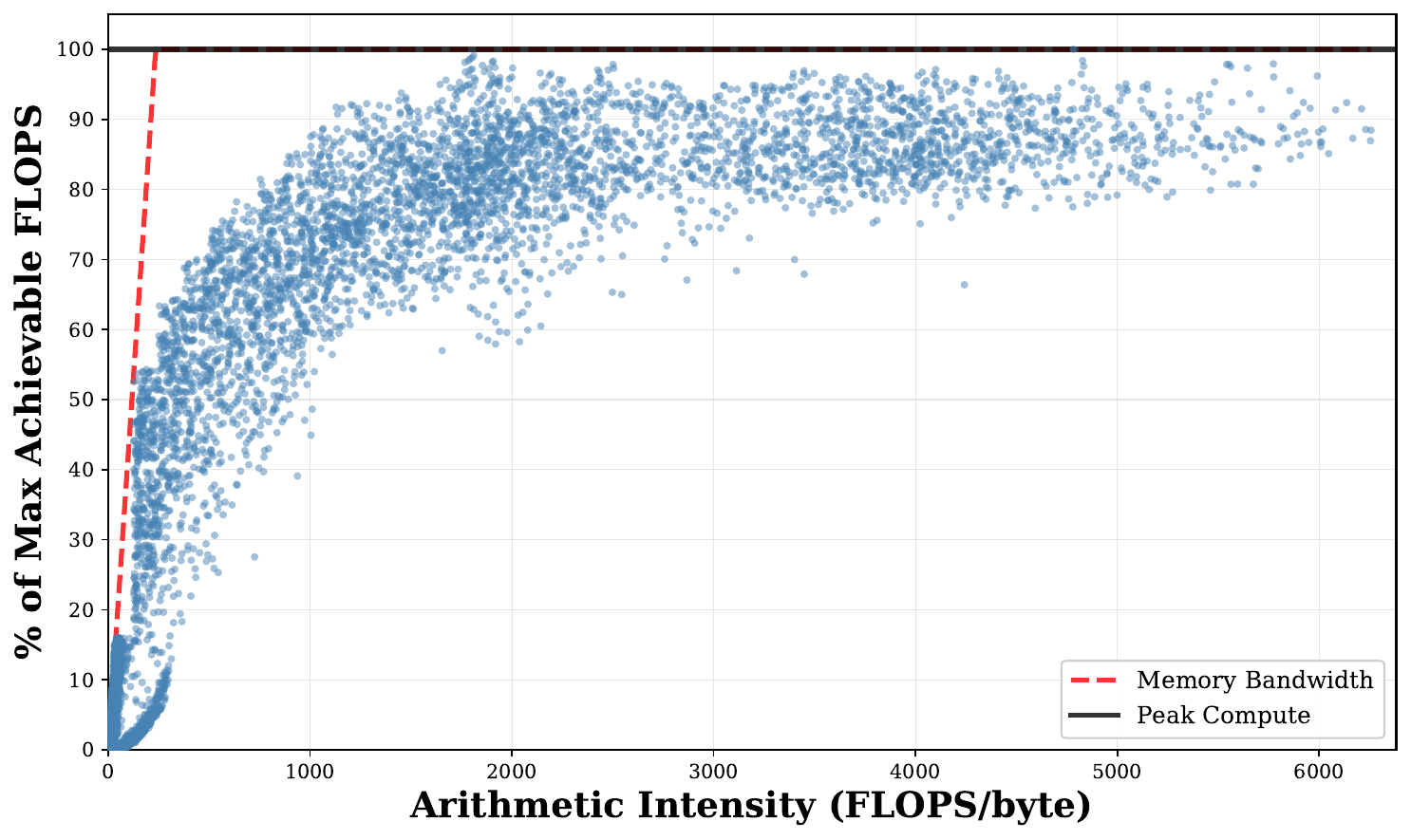}
    \caption{GEMM performance on MI350X for production shapes, shown as percentage of peak achievable FLOPS.}
    \label{fig:origami-roofline}
\end{figure}

\myparagraph{Results.}
Fig.~\ref{fig:origami-roofline} shows Origami performance on production GEMM shapes from recommendation models. The distribution exhibits classical roofline behavior: memory-bound at low arithmetic intensity and compute-bound at high intensity. Compared to exhaustively autotuned GEMM performance, Origami achieves 93\% average selection efficiency across datatypes and transpose configurations. Origami is also integrated into tritonBLAS~\cite{Swann:tta:2025}, enabling zero-warmup kernel selection for arbitrary matrix shapes at runtime, eliminating the need for offline autotuning entirely.

\rebuttal{\myparagraph{Generality.} Although calibrated for MI350X, Origami's formulation is general: because GEMM has regular tiling and predictable data movement, its cost reduces to a hardware-parameterized roofline. Retargeting another accelerator is therefore largely a matter of substituting a handful of hardware constants (cache sizes, achievable bandwidths, matrix-unit throughput, and chiplet topology) rather than re-deriving the model from scratch.}

\subsection{GPU Power Excursion Boost}
\label{sec:power-excursion}

\begin{figure}[t]
    \centerline{\includegraphics[width=0.9\columnwidth]{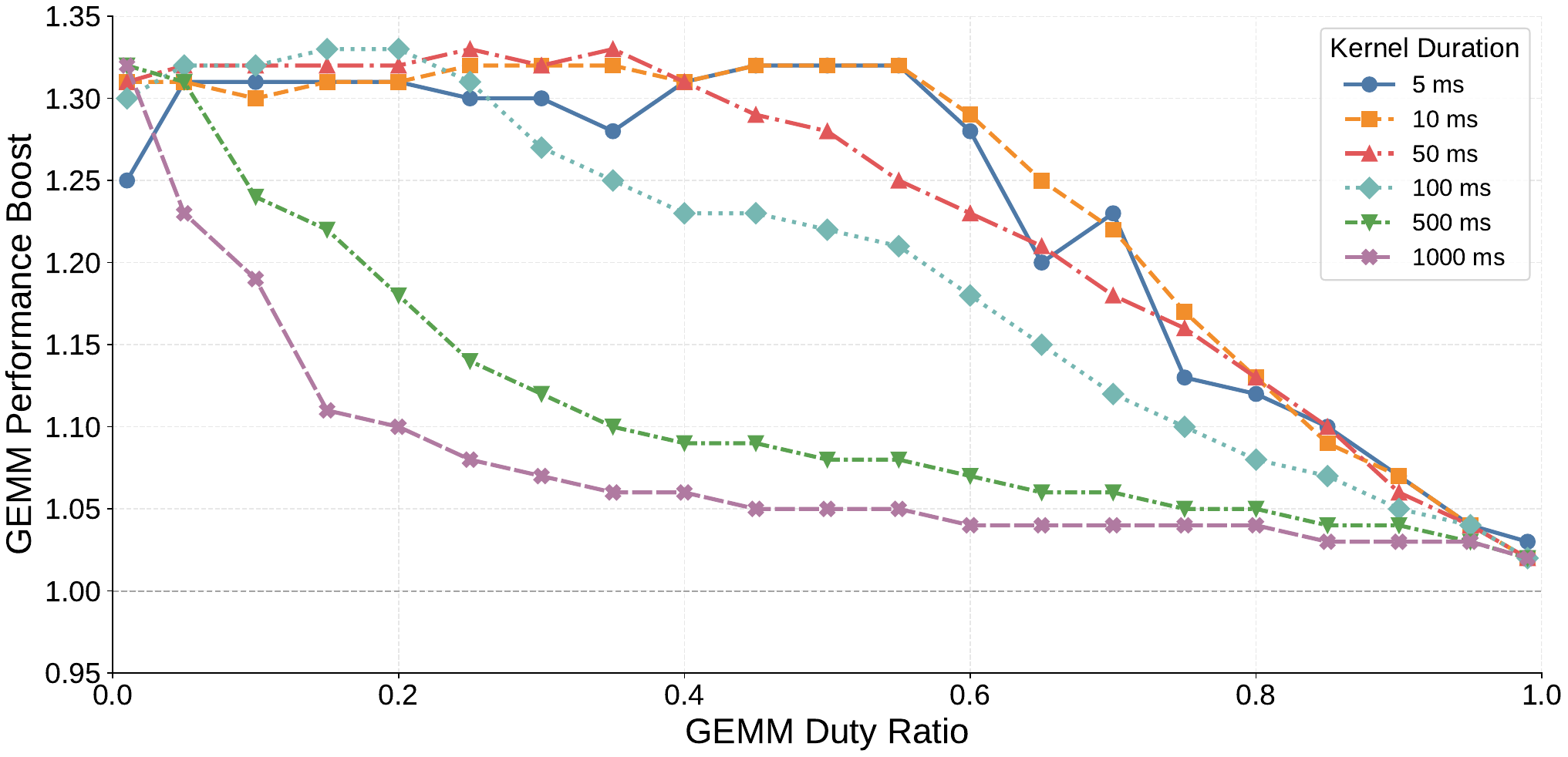}}
    \caption{GEMM performance ($M, N, K$ dimensions $= 8192$) with GPU power excursion across kernel durations and duty ratios on MI350X.}
    \label{fig:power-boost}
\end{figure}

The Open Compute Project (OCP) is a collaborative initiative between hyperscale datacenter operators and technology vendors to standardize interfaces and specifications across infrastructure domains, including power delivery.  The MI350X accelerator is implemented as a 1000\,W open accelerator module (OAM)~\cite{OCP:OAM:2023}. While the OAM must comply with the nominal total board power (TBP) over a 1s window, the specification permits transient power excursions over shorter intervals as defined by the OAM’s excursion design power (EDP). These excursions range from 1.1$\times$~TBP over 20 ms up to 2.0$\times$~TBP over 20 $\mu$s.

Although the OCP specification defines \textit{minimum} electrical, thermal, and power‑delivery capabilities, system designers may provision additional margin. Meta’s AI hardware platform, for example, supports average power levels of up to 1.36$\times$ TBP over a 50\,ms window for a 1000\,W OAM configuration.
By dynamically managing transient power excursions, we leverage the thermal headroom of the MI350X to exceed steady-state power constraints. This allows compute-heavy kernels to surpass the nominal roofline limit, a strategy particularly effective for the hybrid execution profiles of DLRMs, where compute-intensive kernels are interleaved with lower-power operations.

To optimize excursion response under these extended capabilities, we developed a synthetic microbenchmark emulating heterogeneous execution phases. Fig.~\ref{fig:power-boost} measures performance boost by varying the GEMM duty ratio (percentage of high-power GEMM kernels interleaved with lower-power operations such as collectives or elementwise kernels), relative to steady-state operation at 1000\,W with 100\% back-to-back GEMMs. Depending on per-kernel duration, we achieve up to 1.33$\times$ performance gain. This improvement results from tuning the control-loop response to shape the boost profile, maximizing performance over the 50\,ms window across varying phase periods and duty cycles while strictly respecting EDP limits.

In end‑to‑end applications, the benefit from higher excursion limits depends on the frequency and duration of compute‑intensive phases, intervening low‑power operations, and sensitivity to clock frequency.  In Meta's DLRM production training, this enhancement resulted in end‑to‑end performance gains of 3--6\%.

\subsection{Collective Communication with RCCL}

RCCL provides collective communication primitives for AMD GPUs, supporting PCIe, Infinity Fabric Interconnect~\cite{kolla2025xgmi}, and network fabrics (InfiniBand, TCP/IP) across single and multi‑node deployments.

\begin{figure}[t]
    \centerline{\includegraphics[width=0.8\columnwidth]{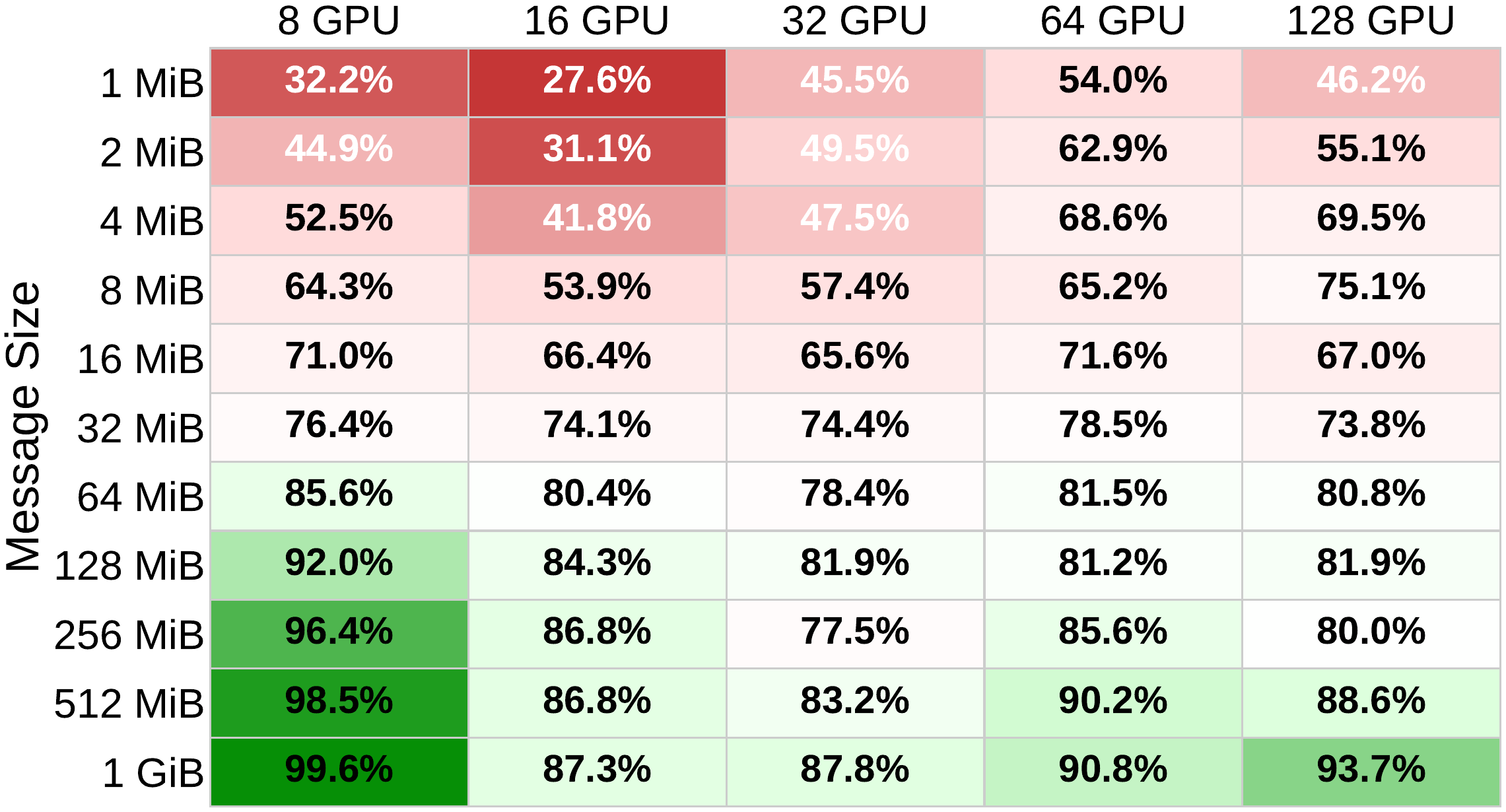}}
    \caption{Measured \texttt{AllToAll} bandwidth with different group and message sizes relative to the theoretical upper bounds.}
    \label{fig:a2a-heatmap}
\end{figure}

We developed a theoretical performance projection model derived from hardware-level simulations of the memory subsystem, scale-up fabric, and NIC data paths\footnote{These projections serve as a theoretical upper bound. In practice, observed bandwidth for small-to-medium messages falls below hardware limits due to software overheads.}. Guided by these projections, we co-designed a suite of RCCL optimizations across the RDMA transport, GPU kernel, and protocol layers for the MI350X. These refinements narrowed the communication gap from $2\times$ slower than NCCL to within 10--15\% of NCCL performance, enabling production-scale DLRM training. Fig.~\ref{fig:a2a-heatmap} shows measured \texttt{AllToAll} bandwidth of different message sizes against performance upper bound.

\myparagraph{Transport and protocol refinements.}
To resolve a performance conflict between collective types, we adopted collective-specific RDMA queue pair (QP) scaling: \texttt{AllToAll} utilizes a single QP for large-chunk efficiency and 30\% higher bus bandwidth, while ring-based collectives (e.g., \texttt{AllGather}) retain four QPs to saturate network bandwidth. We replaced the LL128 low-latency protocol---whose 128-byte stores split across AMD's 64-byte cache lines---with LL64, a protocol that packs 56-byte payload with 8-byte flags into one cache-line-atomic 64-byte packet, delivering $2\times$ \texttt{AllGather}/\texttt{ReduceScatter} and $1.2\times$--$1.4\times$ \texttt{AllReduce} speedups.

\myparagraph{Algorithmic and routing enhancements.}
We introduced Direct \texttt{AllGather} to replace multi-hop ring transfers with simultaneous point-to-point sends, yielding a $1.4\times$ speedup for small messages.
Our P2P batching design reduced \texttt{AllToAll} latency by 40\% at a 128-GPU scale through the aggregation of concurrent communications.

\myparagraph{GPU resource and occupancy optimization.}
A key challenge in production DLRM training is maximizing compute-communication overlap while minimizing contention in shared compute and memory resources. To mitigate CU contention, we transitioned from a thread-block-per-GPU to a warp-per-GPU communication model, capped active channels at 48 (down from 64).
By increasing the thread block size from 256 to 512, we achieved a 46\% VGPR (vector general-purpose register) reduction and doubled occupancy.
These optimizations delivered gains of 63\% for \texttt{ReduceScatter} and 48\% for \texttt{AllReduce}, while reducing the performance penalty when operating with restricted CU allocations.

\subsection{Shampoo Optimizer and rocSOLVER}
\label{sec:rocsolver}

Shampoo~\cite{icml18-shampoo,arxiv23-shampoo} is a second-order optimizer widely deployed due to its faster convergence compared to first-order methods like Adam~\cite{iclr14-adam}. Shampoo maintains Kronecker-factored preconditioner matrices,
requiring periodic computation of matrix fourth roots to update the preconditioners. This matrix root computation relies on \textit{eigen decomposition} provided by the rocSOLVER library.
For large models, these steps dominate Shampoo's overhead, making rocSOLVER's performance critical to overall training throughput.

The \emph{de facto} algorithm to compute the eigen decomposition of a symmetric input matrix $A$ has three steps: (1) reduce $A$ into tridiagonal form, yielding an orthogonal $U$ and a tridiagonal $T$ such that $A = U T U^T$; (2) compute eigenvalues $\Lambda$ and eigenvectors $V$ of $T$; and (3) backtransform eigenvectors $Q = UV$ so that $A = Q \Lambda Q^T$. In rocSOLVER, reduction to tridiagonal form is accomplished with block Householder transformations~\cite{householder1968} operating on panels of the input matrix; updates to the rest of the matrix are applied once per panel as a pair of GEMM operations. Step (2), based on Cuppen’s Divide-and-Conquer (D.C.)~\cite{cuppen1980}, is the most complex step and the focus of our optimizations.

Cuppen’s D.C. splits $T$ into sub-matrices that are processed independently and then merged in pairs. Typically, D.C. operates on unreduced sub-blocks of $T$ (i.e., blocks separated by zero off-diagonal elements), but this can lead to load unbalancing: sub-optimal CU allocation often follows because the size of sub-blocks is input-dependent. Our implementation in rocSOLVER lifts this restriction, yielding precise control over CU allocation for improved performance.
During a merge step, D.C. identifies deflated eigenvalues (which will not be modified by the merge) and updates all others by solving the secular equation. To improve data locality and reduce data movement, our implementation ensures that eigenvalues (and associated eigenvectors) are properly ordered. Eigenvectors belonging to identical eigenvalues require (possibly costly) extra preprocessing; rocSOLVER includes specialized code to efficiently handle matrices exhibiting several identical eigenvalues in merge steps. Lastly, newly and previously computed eigenvectors are merged using batched GEMMs for optimal performance.

\begin{figure}[t]
    \centerline{\includegraphics[width=0.7\columnwidth]{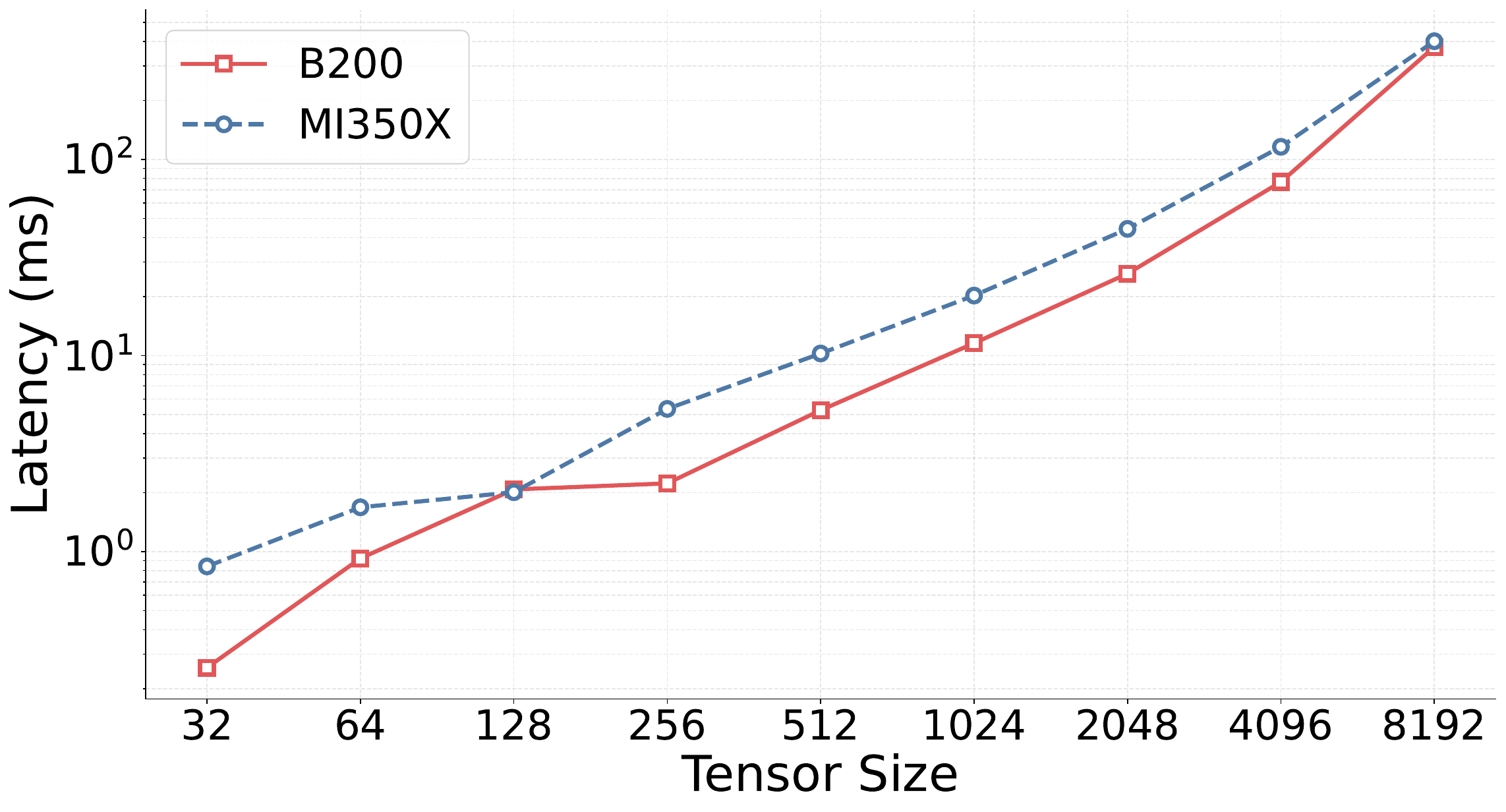}}
    \caption{Eigen decomposition performance.}
    \label{fig:eigensolver}
\end{figure}

Fig.~\ref{fig:eigensolver} demonstrates that rocSOLVER achieves comparable eigen decomposition performance on AMD MI350X relative to NVIDIA B200. Distributed Shampoo partitions large preconditioner matrices into fixed-size blocks (typically $4096 \times 4096$ or $8192 \times 8192$) to bound per-GPU computation and enable parallel eigen decomposition across devices. These block sizes represent the practical sweet spot: large enough to preserve preconditioning quality, yet small enough for tractable eigen decomposition. rocSOLVER narrows the gap with cuSOLVER at these critical 4K and 8K dimensions, enabling competitive Shampoo-based training on AMD hardware.

\section{Kernel Tuning and Agentic AI Workflows}
\label{sec:kernel}

In this section, we first identify kernel optimization principles through case studies of key production kernels, and then encode them as reusable skills in an agentic AI workflow that scale to the long tail of model-specific operators.

\subsection{Triton Kernel Optimizations}
\label{sec:triton-kernels}

\subsubsection{Attention Kernels in DLRM} We focus on two important Triton kernels in Meta's production DLRM workloads, GDPA and HSTU, and describe the AMD-specific optimization techniques that yield substantial speedups on MI350X.

\myparagraph{Generalized dot-product attention (GDPA).}
GDPA~\cite{pytorch26-gdpa} extends standard attention~\cite{neurips17-attn, neurips22-flash-attn} by fusing GeLU-activated attention and a feed-forward network into a single operator.
Central to Meta's Kunlun~\cite{arxiv26-kunlun} and Interformer~\cite{arxiv25-interformer} models, the fused kernel substantially increases the compute and memory intensity, positioning the kernel as a critical optimization target.

\myparagraph{Hierarchical sequential transduction unit (HSTU).}
HSTU~\cite{icml24-hstu} is a novel sequential attention architecture designed for large-scale recommendation systems, where the input consists of variable-length user interaction histories with high sparsity.
Unlike standard attention, HSTU uses variable-length batching and processes sequences with sparsity ratios as high as 0.95, meaning the effective computation is concentrated in a small fraction of the sequence.

\begin{figure*}[t]
  \centering
  \begin{subfigure}[t]{0.24\textwidth}
    \centering
    \includegraphics[width=\linewidth]{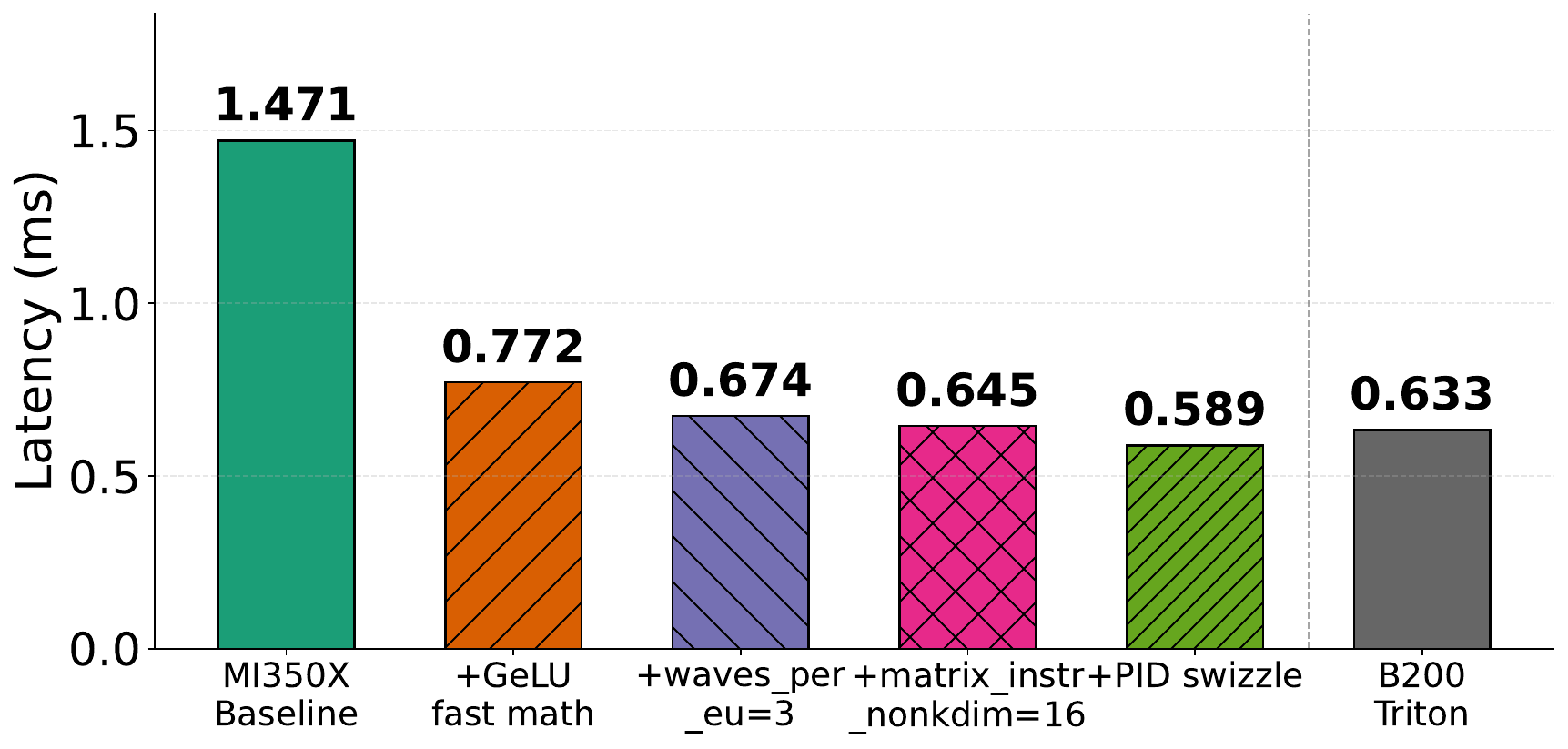}
    \caption{Forward latency (ms).}
    \label{fig:gdpa-fwd-latency}
  \end{subfigure}\hfill
  \begin{subfigure}[t]{0.24\textwidth}
    \centering
    \includegraphics[width=\linewidth]{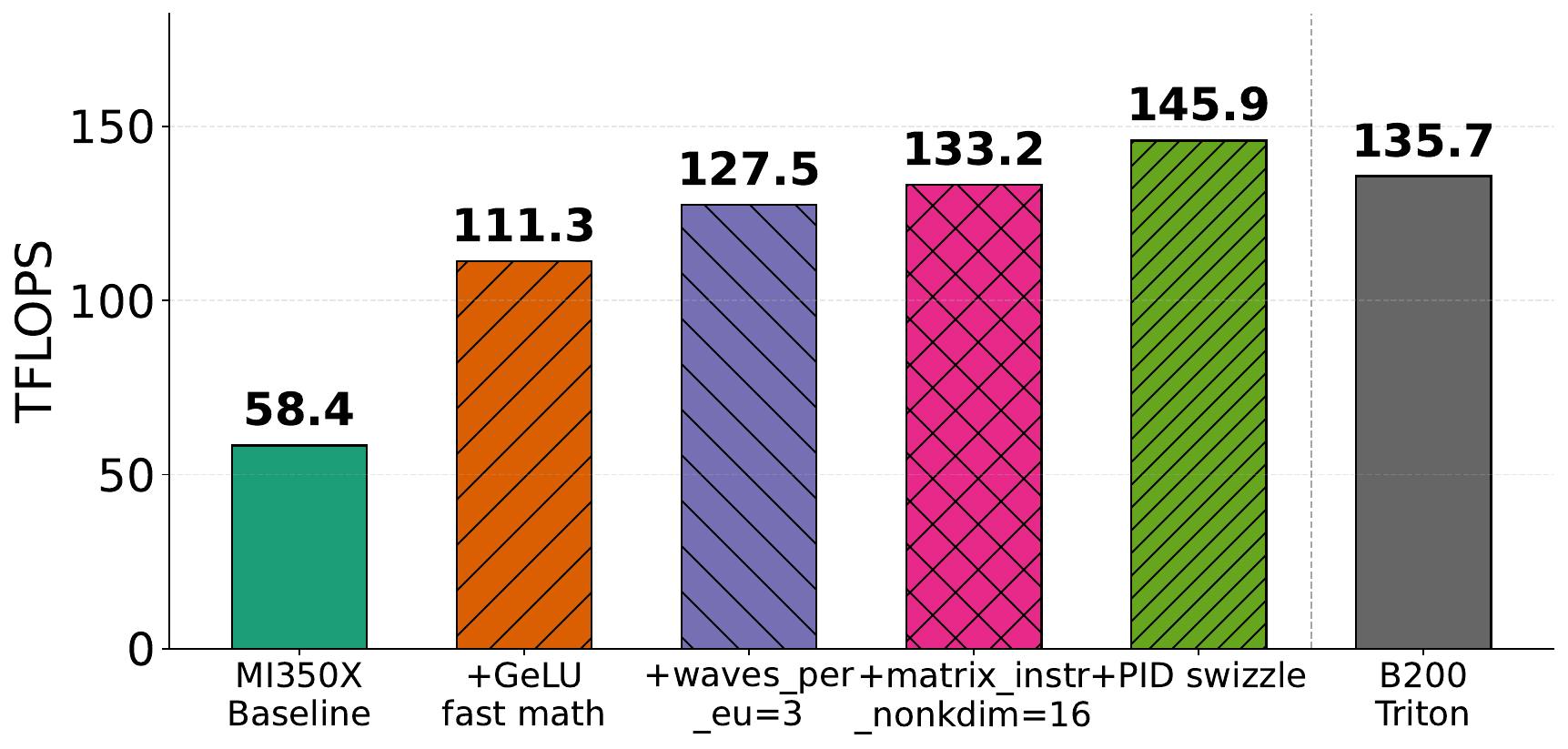}
    \caption{Forward throughput (TFLOP/s).}
    \label{fig:gdpa-fwd-tflops}
  \end{subfigure}\hfill
  \begin{subfigure}[t]{0.24\textwidth}
    \centering
    \includegraphics[width=\linewidth]{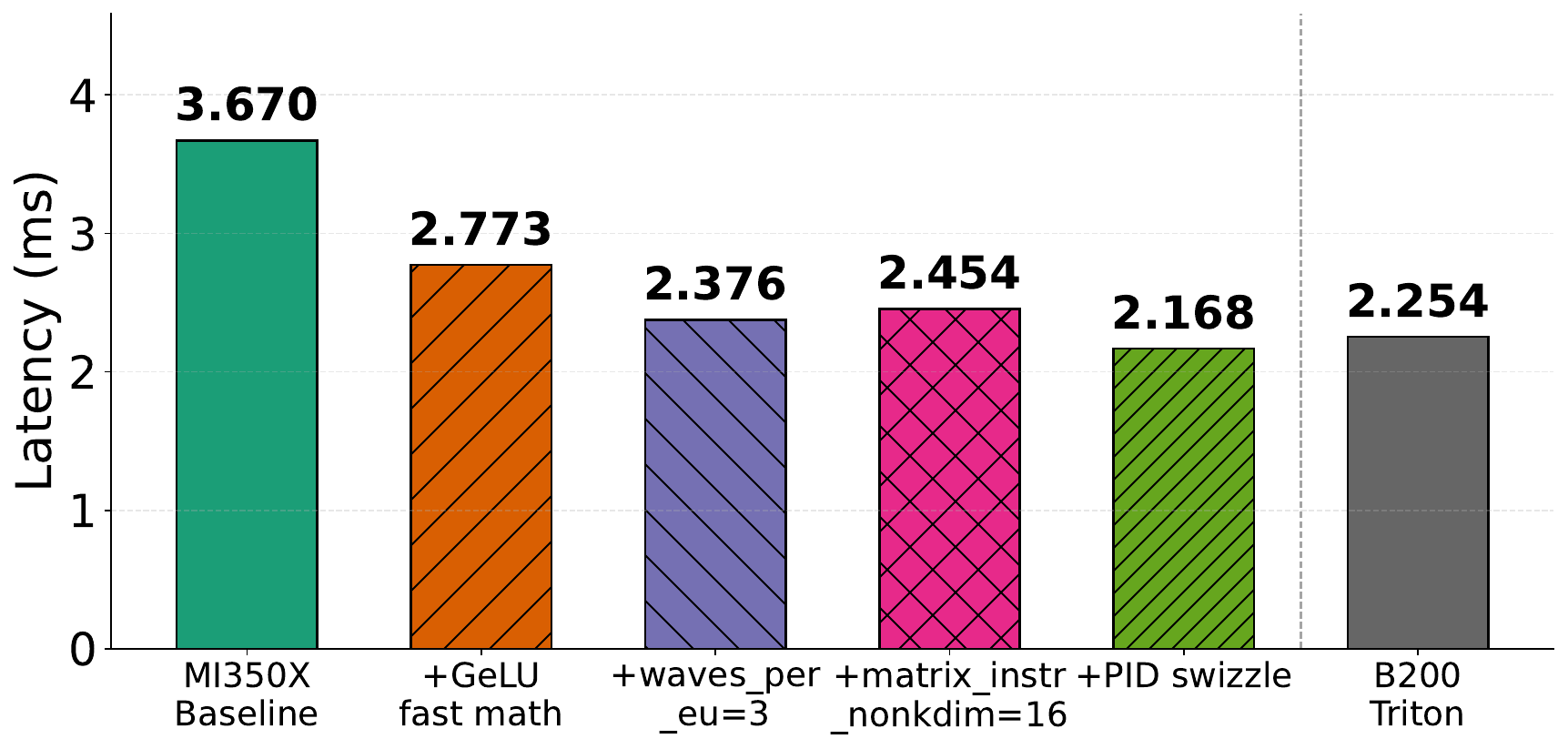}
    \caption{Backward latency (ms).}
    \label{fig:gdpa-bwd-latency}
  \end{subfigure}\hfill
  \begin{subfigure}[t]{0.24\textwidth}
    \centering
    \includegraphics[width=\linewidth]{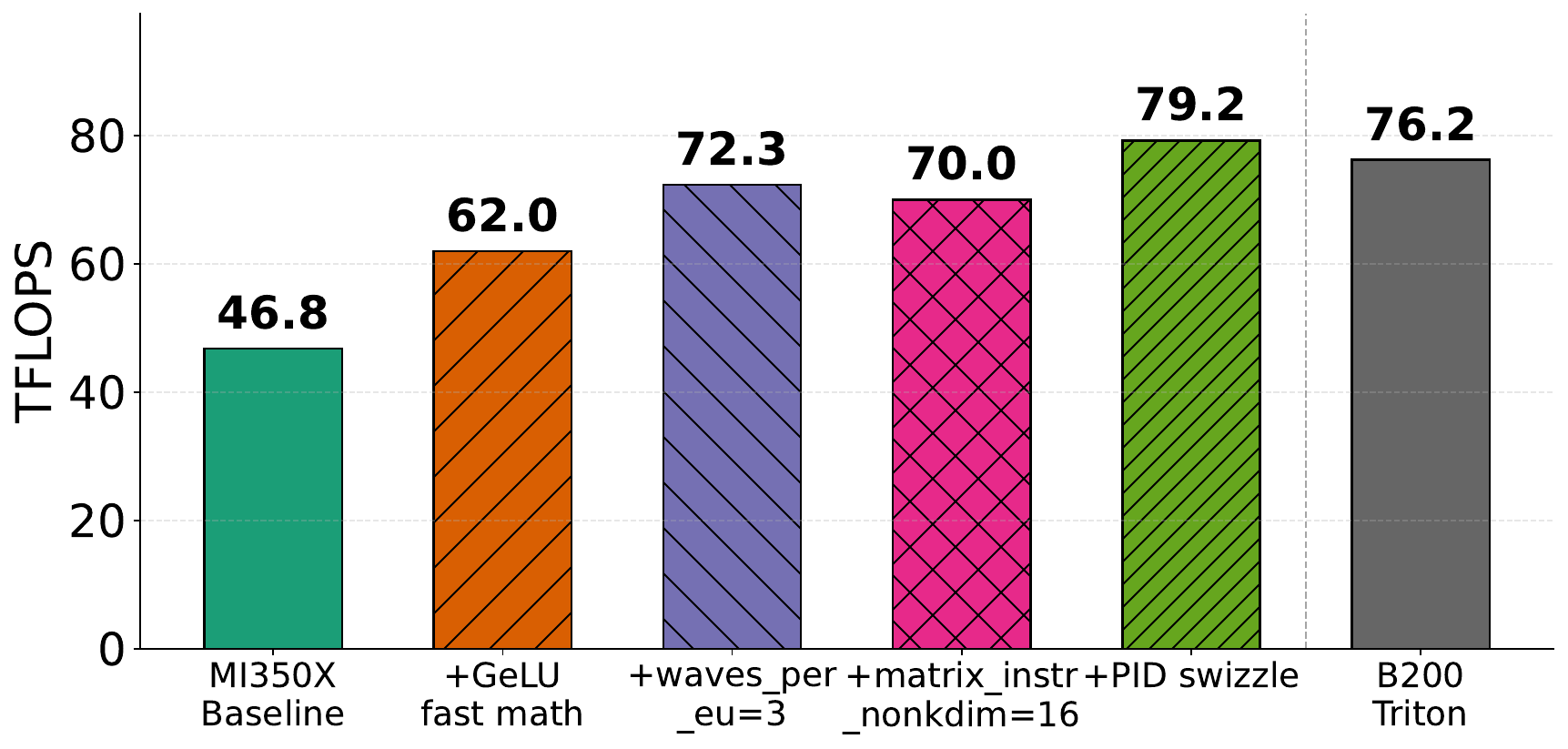}
    \caption{Bwd throughput (TFLOP/s).}
    \label{fig:gdpa-bwd-tflops}
  \end{subfigure}
  \caption{GDPA kernel performance with incremental optimization techniques applied on AMD MI350X, compared to optimal configuration on B200. Lower latency and higher throughput are better.}
  \label{fig:gdpa-results}
  \vspace{-5pt}
\end{figure*}

\begin{figure*}[t]
  \centering
  \begin{subfigure}[t]{0.24\textwidth}
    \centering
    \includegraphics[width=\linewidth]{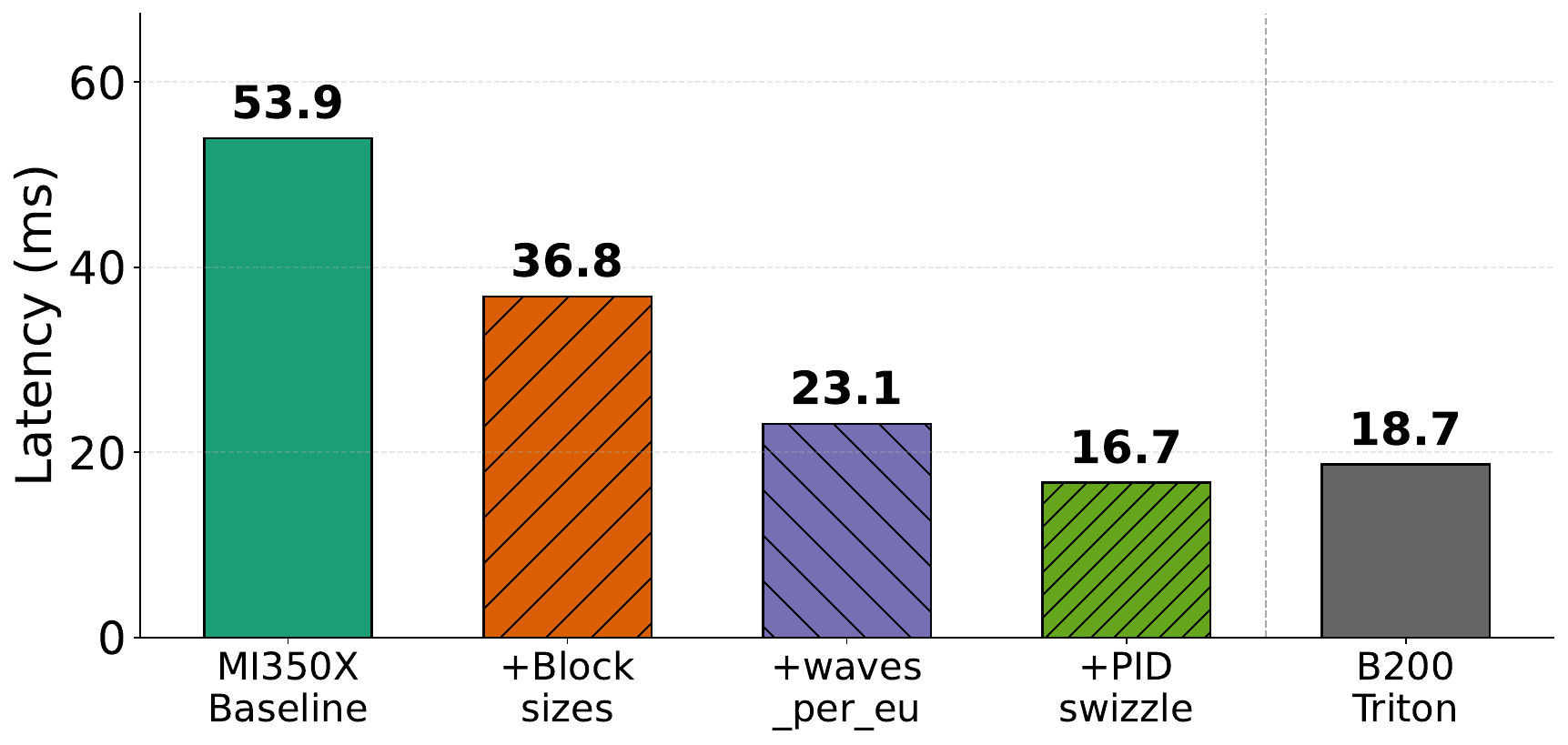}
    \caption{Forward latency (ms).}
    \label{fig:hstu-fwd-latency}
  \end{subfigure}\hfill
  \begin{subfigure}[t]{0.24\textwidth}
    \centering
    \includegraphics[width=\linewidth]{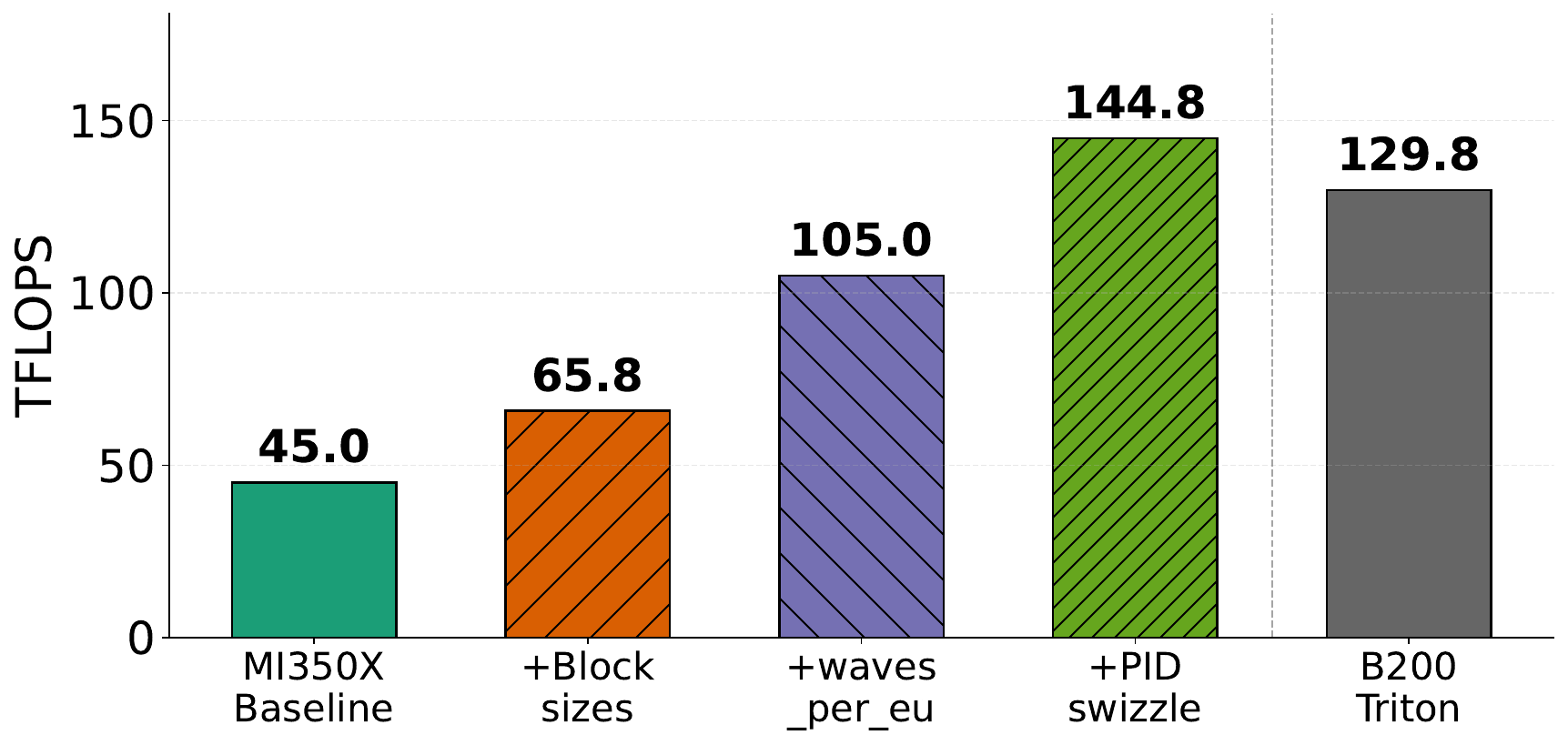}
    \caption{Forward throughput (TFLOP/s).}
    \label{fig:hstu-fwd-tflops}
  \end{subfigure}\hfill
  \begin{subfigure}[t]{0.24\textwidth}
    \centering
    \includegraphics[width=\linewidth]{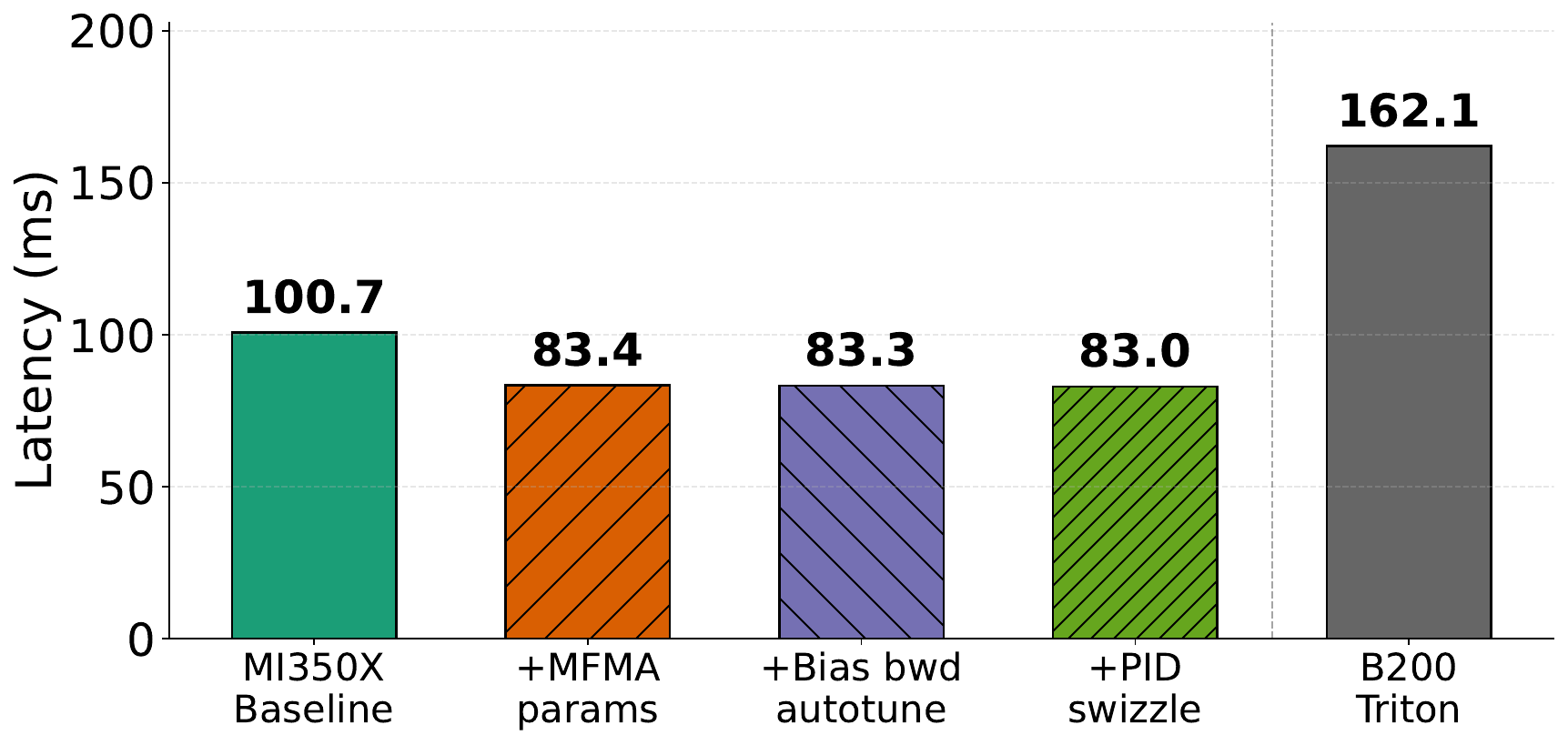}
    \caption{Backward latency (ms).}
    \label{fig:hstu-bwd-latency}
  \end{subfigure}\hfill
  \begin{subfigure}[t]{0.24\textwidth}
    \centering
    \includegraphics[width=\linewidth]{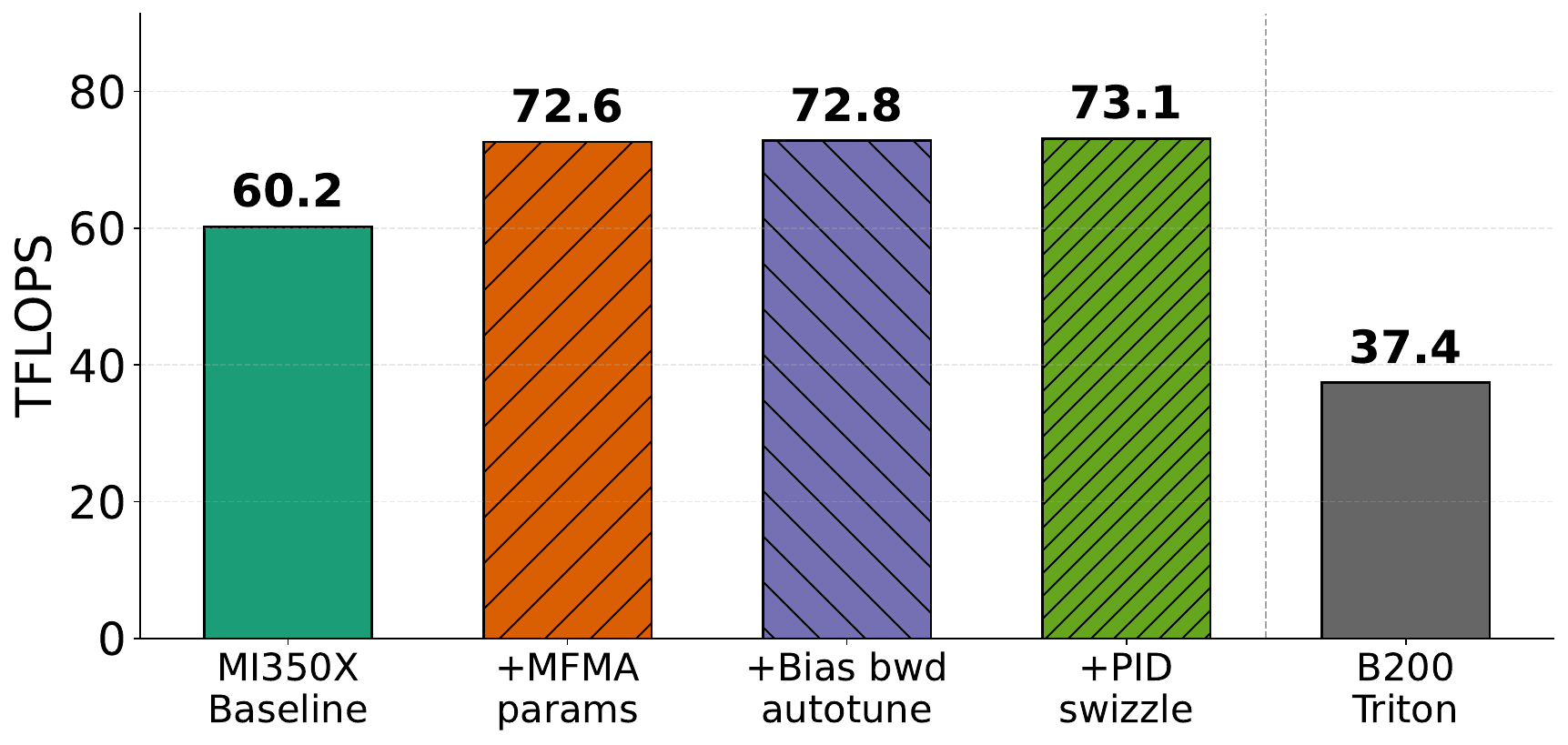}
    \caption{Bwd throughput (TFLOP/s).}
    \label{fig:hstu-bwd-tflops}
  \end{subfigure}
  \caption{HSTU kernel performance with incremental optimization techniques applied on AMD MI350X, compared to optimal configuration on B200. Lower latency and higher throughput are better.}
  \label{fig:hstu-results}
  \vspace{-10pt}
\end{figure*}

\subsubsection{Optimization Techniques}

We apply five optimization techniques to the Triton kernels, each addressing a different performance bottleneck on the MI350X architecture.
While we use GDPA and HSTU as examples, these optimization techniques are generally applicable to a broad set of Triton kernels for ML models.

\myparagraph{PID swizzle.}
MI350X features a multi-XCD (chiplet) architecture where eight XCDs share an L2 cache.
The default linear mapping of Triton program IDs to tiles can cause XCDs to access disjoint L2 cache regions, resulting in poor cache utilization.
PID swizzle uses \texttt{tl.swizzle2d} to remap the two-dimensional program ID grid so that spatially adjacent tiles are scheduled on the same XCD, improving L2 cache locality.

\myparagraph{Occupancy hints.}
The \texttt{waves\_per\_eu} hint controls the maximum number of wavefronts resident on each execution unit, which directly affects register allocation.
By setting \texttt{waves\_per\_eu=3} (GDPA) or exposing it as an autotunable parameter with values 0--4 (HSTU), the compiler reduces per-wavefront register usage, allowing more concurrent wavefronts to hide memory latency.
This trades per-wavefront performance for higher occupancy, which is beneficial when the kernel is not fully compute-bound.

\myparagraph{MFMA instruction configuration.}
AMD CDNA architectures provide matrix fused multiply-add (MFMA) instructions in configurable tile sizes, controlled by the \texttt{matrix\_instr\_nonkdim} parameter.
By setting \texttt{matrix\_instr\_nonkdim=16}, the system selects $16\times16$ MFMA tiles, which on CDNA~4 can achieve higher instruction throughput than the default $32\times32$ tiles for certain matrix shapes.


\myparagraph{Triton configuration autotuning.}
Triton kernel configurations (\texttt{BLOCK\_SIZE}, \texttt{num\_stages}, \texttt{num\_warps}, etc.) are critical to performance, and optimal values differ between NVIDIA and AMD GPUs due to architectural differences.
AMD's 64-thread wavefronts, compared to NVIDIA's 32-thread warps, favor larger \texttt{num\_warps} values to achieve sufficient occupancy and hide memory latency.
NVIDIA's Tensor Memory Accelerator (TMA) on H100/B200 enables deep software pipelining (e.g., \texttt{num\_stages} $\in[4, 7]$) by efficiently managing asynchronous memory transfers.
AMD MI350X lacks a TMA equivalent; high stage counts increase LDS pressure without commensurate latency hiding benefits.
Empirically, AMD kernels achieve better performance with shallower pipelines (e.g., \texttt{num\_stages} $\approx 2$).

\myparagraph{GeLU fast math.}
The default \texttt{fast\_gelu} implementation in Triton uses a \texttt{tanh}-based approximation, which on AMD GPUs decomposes into $\tanh(x) = 2\sigma(2x) - 1$, requiring \rebuttal{a sigmoid evaluation (scaled by two)} and additional arithmetic.
We replace this with a direct sigmoid-based GeLU approximation using the compiler intrinsics \texttt{fast\_expf} and \texttt{fast\_dividef}, which map to single hardware instructions with better performance.

\subsubsection{Performance Results}

We evaluate these tuning techniques by incrementally applying them to the GDPA and HSTU forward and backward kernels.

\myparagraph{GDPA.}
Fig.~\ref{fig:gdpa-fwd-latency} and~\ref{fig:gdpa-fwd-tflops} show the incremental forward kernel improvements based on a representative configuration\footnote{The GDPA configuration used in this experiment: \texttt{batch=2048, d\_model=256, num\_head=4, head\_dim=64, d\_ff=256, seq=500, sparsity=0.68}.} in GEM models~\cite{meta25-gem}.
Starting from the baseline of 1.47\,ms latency (58.4\,TFLOP/s), the cumulative application of GeLU fast math, occupancy hints, MFMA instruction configuration, and PID swizzle reduces latency to 0.59\,ms (145.9\,TFLOP/s), a total 60.0\% reduction in latency and $2.5\times$ in throughput.
Fig.~\ref{fig:gdpa-bwd-latency} and~\ref{fig:gdpa-bwd-tflops} show the backward kernel results.
The same techniques (except \texttt{matrix\_instr\_nonkdim}, which regresses the backward pass) reduce latency from 3.67\,ms to 2.17\,ms, a 40.9\% reduction.
Compared to B200, the optimized GDPA kernels on MI350X achieve better performance.

\noindent \textbf{HSTU.}
Fig.~\ref{fig:hstu-fwd-latency} and~\ref{fig:hstu-fwd-tflops} show the forward kernel results for a representative configuration\footnote{The HSTU configuration used in this experiment: \texttt{batch=512, num\_head=4, attn\_dim=128, hidden\_dim=128, target\_size=20, sparsity=0.95}. The figures plot \texttt{seq=4096}; averaged results sweep \texttt{seq} over 256, 512, 1024, 2048, and 4096.} in production models.
Block size tuning, occupancy hints, and PID swizzle together yield a $3.22\times$ speedup, with peak throughput improving from 45.0 to 144.8\,TFLOP/s.
Fig.~\ref{fig:hstu-bwd-latency} and~\ref{fig:hstu-bwd-tflops} show the backward kernel results.
Out-of-the-box performance is already stronger on MI350X here, leaving less optimization headroom: MFMA parameter tuning delivers a $1.21\times$ speedup, with peak throughput improving from 60.2 to 72.6\,TFLOP/s.
Averaged across sequence lengths, the optimized MI350X kernels are $1.22\times$ faster than B200 on the forward pass and $1.80\times$ faster on the backward pass, for a combined $1.69\times$ advantage that widens at longer sequences.

\subsection{CUDA Kernel Optimizations}
\label{sec:cuda-kernels}

GPU kernels written in CUDA can be compiled for AMD GPUs via HIP translation (Section~\ref{sec:bg-amd}).
While it enables code portability, certain CUDA idioms, particularly those relying on atomic operations, perform poorly on AMD's multi-chiplet architecture.
This section describes the architectural differences that cause these performance gaps and our algorithmic strategies to address them.

\subsubsection{Atomic Operations on Multi-Chiplet GPUs}

Unlike NVIDIA GPUs where atomic operations complete in a monolithic L2 cache, MI350X atomics may require cross-chiplet cache coherency (note 8 XCD chiplets in MI350X), incurring round-trip latencies across the Infinity Fabric.
When thousands of threads contend on a single address (as is common in reduction-style kernels), these coherency messages serialize execution, leading to poor performance.

The fundamental issue is that per-thread global atomics scale poorly with thread count on multi-chiplet architectures.
On a monolithic GPU, the L2 cache can coalesce nearby atomic operations; on a multi-chiplet GPU, each chiplet must coordinate with the others, and a single contended address becomes a serialization bottleneck.
Our measurements show that MI350X can be over $10\times$ slower than B200 on atomic-heavy kernels at high contention rates, while matching or exceeding B200 on atomic-free, bandwidth-bound kernels.

\noindent \textbf{Algorithmic strategies: hierarchical reduction and caching.}
We employ two complementary strategies depending on the contention pattern. When many threads contend on a \emph{single shared address} (e.g., a global counter), we apply \emph{hierarchical reduction}: threads accumulate results in fast on-chip memory (registers or shared memory), a parallel tree reduction sums them within each block, and a single representative thread issues one global atomic. This converts $N$ global atomics into $\lceil N / \text{block\_size} \rceil$, a reduction factor of up to $1024\times$. When threads target many distinct addresses with temporal locality (e.g., gradient scatter to embedding rows), we apply \emph{caching}: each thread maintains a small register cache of recently written addresses, merging consecutive updates to the same address locally before issuing a single atomic add on eviction.
The tradeoff is additional shared memory usage. For kernels where contention is rare, the overhead may slightly increase latency, but the benefit grows superlinearly with contention rate, making this transformation highly effective for production workloads with moderate-to-high atomic traffic.

\begin{figure}[t]
  \centering
  \includegraphics[width=0.8\linewidth]{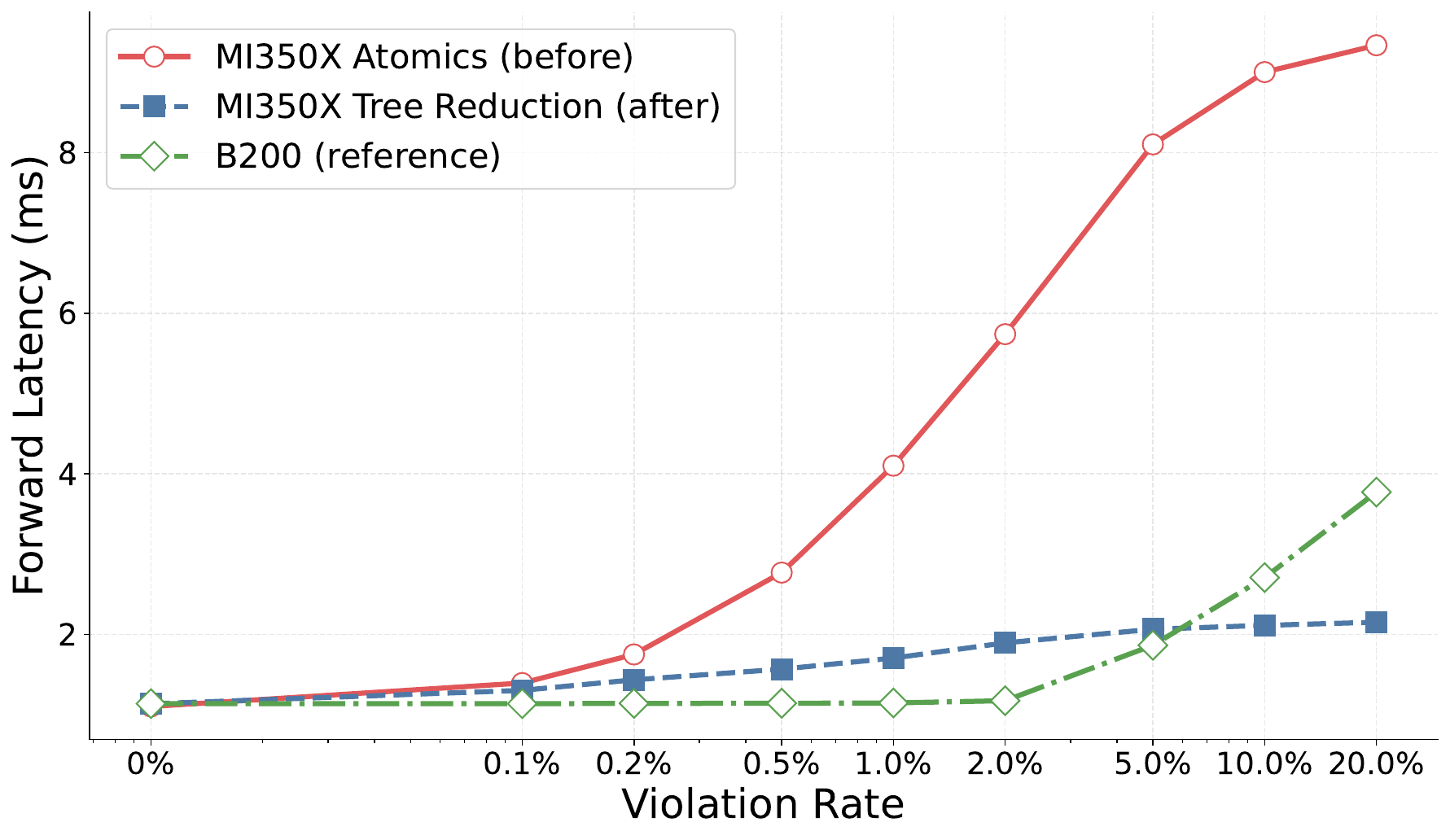}
  \caption{Bounds check forward kernel latency vs.\ violation rate. Tree reduction eliminates atomic contention, enabling MI350X to outperform B200 at high violation rates.}
  \label{fig:bounds-check-fwd}
  \vspace{-5pt}
\end{figure}

\subsubsection{Example: Bounds Check Kernel}

The kernel validates that all embedding table indices fall within valid ranges before performing lookups. When operating in \texttt{WARNING} mode, the kernel (1) accumulates a per-thread count of out-of-bounds violations, (2) corrects each invalid index, and (3) increments the global out-of-bounds violation counter via atomic add operations. On MI350X, atomic operations become severe bottlenecks at high violation rates.

\noindent \textbf{Original implementation.}
After each thread accumulates its violation count, threads with non-zero counts issue \texttt{gpuAtomicAdd}s to global memory. With 1024 threads per block and thousands of blocks processing millions of indices, a typical workload can produce millions of threads that contend on a single address.

\noindent \textbf{Optimized implementation.}
In order to reduce atomic operations, per-thread global adds are replaced with shared-memory parallel tree reduction followed by a single atomic add. Rather than writing to a global counter directly, each thread writes its local violation count to a shared memory buffer allocated at the block level. A 10-step binary reduction tree then sums the per-thread contributions entirely within shared memory, avoiding any cross-XCD traffic. Finally, thread 0 issues a single \texttt{gpuAtomicAdd} with the block-wide sum to the global counter, reducing global atomics from \scalebox{0.8}[1]{$\sim$}$10^6$ to \scalebox{0.8}[1]{$\sim$}$10^3$ per batch for a typical production workload.

\noindent \textbf{Results.}
Fig.~\ref{fig:bounds-check-fwd} shows the latency on a production embedding configuration (30 tables, 13.8M indices) as violation rate varies from 0\% to 20\%.
With per-thread atomic adds, MI350X forward latency degrades $8.5\times$ from 0\% to 20\% violations while B200 degrades only $3.3\times$.
At 10\% violations, tree reduction delivers $4.26\times$ speedup (2.1\,ms vs 9.0\,ms).
At 20\% violations, the speedup reaches $4.34\times$, and MI350X with tree reduction is $1.75\times$ faster than B200.

\begin{figure}[t]
  \centering
  \begin{subfigure}[t]{0.235\textwidth}
    \centering
    \includegraphics[width=\linewidth]{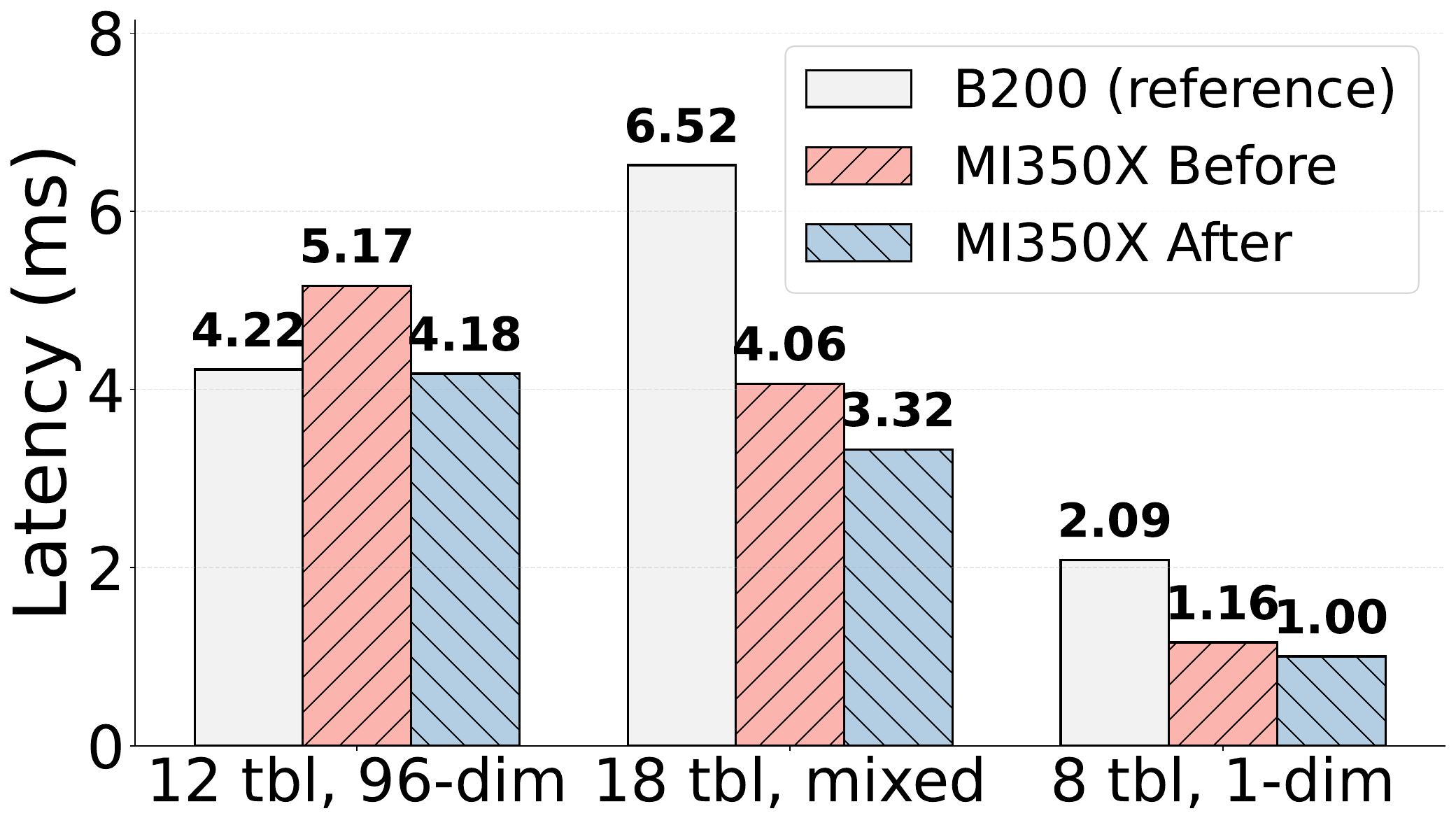}
    \caption{Forward latency (ms).}
    \label{fig:gis-fwd}
  \end{subfigure}\hfill
  \begin{subfigure}[t]{0.235\textwidth}
    \centering
    \includegraphics[width=\linewidth]{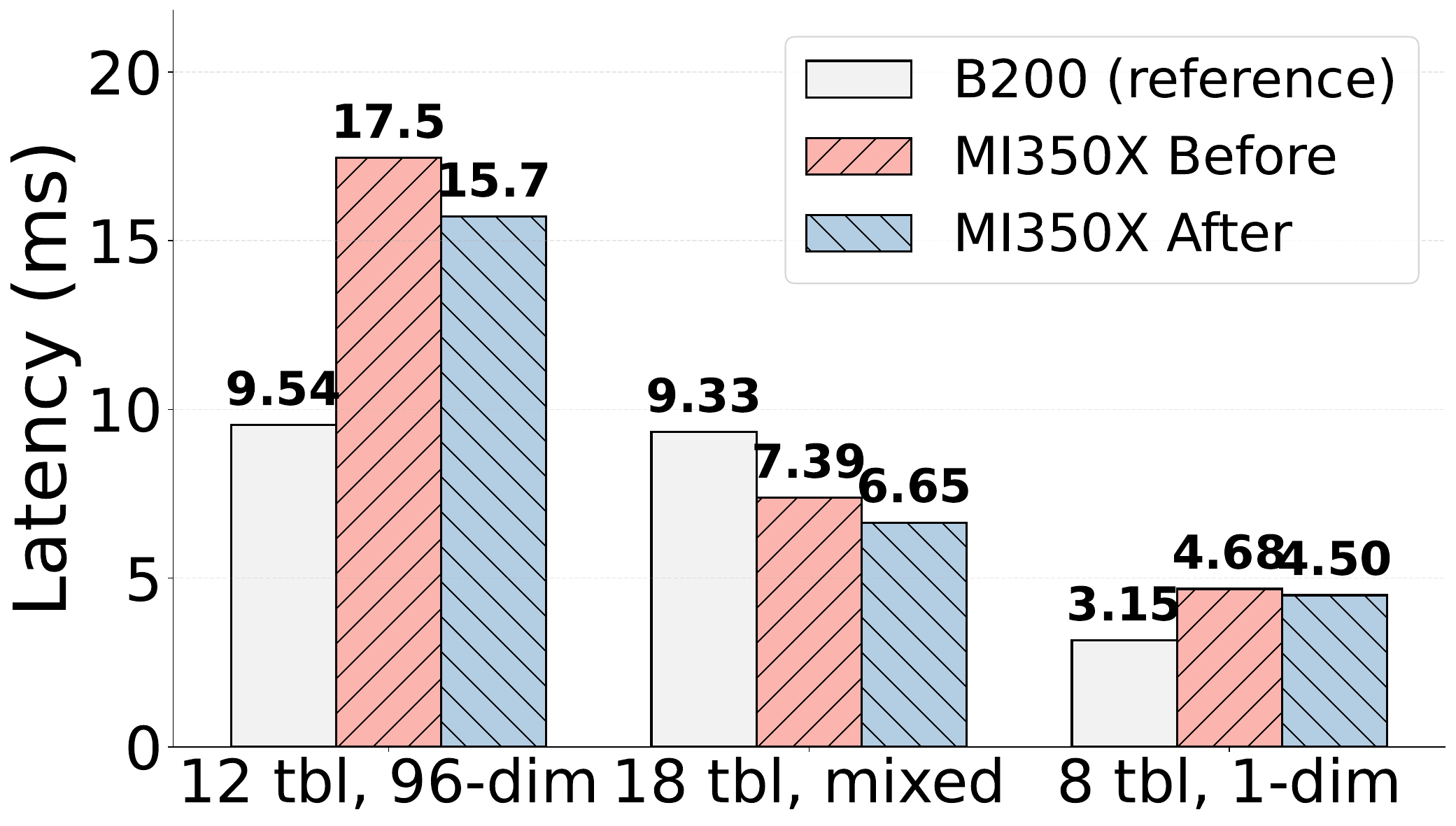}
    \caption{Backward latency (ms).}
    \label{fig:gis-bwd}
  \end{subfigure}
  \caption{GIS kernel performance. For the bandwidth-bound forward kernel, MI350X exceeds B200. For backward, multi-row packing helps on mixed-dimension configs but cannot fully overcome atomic contention on uniform, high-collision workloads.}
  \label{fig:fbgemm-gis}
\end{figure}

\subsubsection{Example: Grouped Index Select (GIS) Kernel}
A core operation in recommendation model training, GIS gathers rows across multiple embedding tables in a single fused kernel. The forward pass is a bandwidth-bound simple gather, while the backward pass scatters gradients back via atomic add, with contention arising when multiple samples select the same row.

\noindent \textbf{Forward optimization.}
AMD's native 64-thread wavefronts underutilize hardware when embedding dimensions are small. Thus, we introduce 32-thread emulated warps that enable multiple rows to be packed into the same wavefront, eliminating idle lanes and improving memory bandwidth utilization. This optimization delivers $1.16\times$--$1.24\times$ speedup across production configurations (Fig.~\ref{fig:gis-fwd}).

\noindent \textbf{Backward optimization.}
In the backward pass, we exploit \emph{temporal locality} within each thread's iteration stream. Each thread maintains a 2-slot register cache that maps row indices to accumulated gradient values. When consecutive iterations target the same row (a cache hit), gradients are summed locally in registers with zero global memory traffic. On a miss with both slots occupied, the oldest entry is evicted via a single \texttt{atomicAdd} and replaced. This converts $N$ consecutive updates to the same row into one atomic, with the coalescing factor determined by the locality of the index stream, delivering $1.04\times$--$1.11\times$ backward speedup (Fig.~\ref{fig:gis-bwd}).


\begin{table}[t]
  \footnotesize
  \setlength{\tabcolsep}{2pt}  
  \definecolor{deepgreenbg}{HTML}{4CAF50}
  \definecolor{greenbg}{HTML}{A5D6A7}
  \definecolor{lightgreenbg}{HTML}{E8F5E9}
  \definecolor{lightredbg}{HTML}{FFEBEE}
  \definecolor{redbg}{HTML}{EF9A9A}
  \definecolor{deepredbg}{HTML}{E57373}

  \begin{tabular}{@{}ccccccc|cc|ccc@{}}
    \toprule
    & & & & & & & \multicolumn{2}{c|}{B200} & \multicolumn{3}{c}{MI350X} \\
    B & T & L & D & Prec & $\alpha$ & & FWD & BWD & FWD & \multicolumn{2}{c}{BWD} \\
    & & & & & & & (ms) & (ms) & (ms) & Before & After \\
    \midrule
    65536 & 5 & 55 & 192 & fp16 & 1.0 &
      & 1.75 & 14.40
      & \cellcolor{greenbg}\makecell{1.48\\(0.85$\times$)}
      & \cellcolor{deepredbg}\makecell{23.6\\(1.64$\times$)}
      & \cellcolor{greenbg}\makecell{7.72\\(0.54$\times$)} \\
    65536 & 5 & 55 & 192 & fp16 & 1.15 &
      & 0.84 & 4.61
      & \cellcolor{lightgreenbg}\makecell{0.78\\(0.93$\times$)}
      & \cellcolor{deepredbg}\makecell{9.92\\(2.15$\times$)}
      & \cellcolor{greenbg}\makecell{3.97\\(0.86$\times$)} \\
    65536 & 10 & 26 & 256 & fp16 & 1.0 &
      & 1.86 & 15.79
      & \cellcolor{lightgreenbg}\makecell{1.84\\(0.99$\times$)}
      & \cellcolor{deepredbg}\makecell{24.4\\(1.55$\times$)}
      & \cellcolor{greenbg}\makecell{8.06\\(0.51$\times$)} \\
    65536 & 10 & 26 & 256 & fp16 & 1.15 &
      & 0.87 & 4.85
      & \cellcolor{lightredbg}\makecell{0.91\\(1.05$\times$)}
      & \cellcolor{deepredbg}\makecell{11.4\\(2.35$\times$)}
      & \cellcolor{greenbg}\makecell{4.36\\(0.90$\times$)} \\
    65536 & 3 & 55 & 192 & fp32 & 1.0 &
      & 1.36 & 7.56
      & \cellcolor{redbg}\makecell{1.50\\(1.10$\times$)}
      & \cellcolor{redbg}\makecell{10.6\\(1.40$\times$)}
      & \cellcolor{greenbg}\makecell{5.60\\(0.74$\times$)} \\
    65536 & 3 & 55 & 192 & fp32 & 1.15 &
      & 0.53 & 2.67
      & \cellcolor{lightgreenbg}\makecell{0.53\\(1.00$\times$)}
      & \cellcolor{deepredbg}\makecell{7.45\\(2.79$\times$)}
      & \cellcolor{lightredbg}\makecell{2.69\\(1.01$\times$)} \\
    65536 & 6 & 26 & 256 & fp32 & 1.0 &
      & 2.29 & 8.23
      & \cellcolor{greenbg}\makecell{2.03\\(0.89$\times$)}
      & \cellcolor{redbg}\makecell{10.9\\(1.32$\times$)}
      & \cellcolor{greenbg}\makecell{5.80\\(0.70$\times$)} \\
    65536 & 6 & 26 & 256 & fp32 & 1.15 &
      & 0.59 & 2.98
      & \cellcolor{redbg}\makecell{0.66\\(1.12$\times$)}
      & \cellcolor{deepredbg}\makecell{8.95\\(3.00$\times$)}
      & \cellcolor{redbg}\makecell{3.60\\(1.21$\times$)} \\
    131072 & 3 & 20 & 256 & fp16 & 1.0 &
      & 0.84 & 6.66
      & \cellcolor{lightredbg}\makecell{0.85\\(1.01$\times$)}
      & \cellcolor{deepredbg}\makecell{10.9\\(1.64$\times$)}
      & \cellcolor{greenbg}\makecell{3.69\\(0.55$\times$)} \\
    131072 & 3 & 20 & 256 & fp16 & 1.15 &
      & 0.43 & 2.39
      & \cellcolor{lightredbg}\makecell{0.46\\(1.07$\times$)}
      & \cellcolor{deepredbg}\makecell{8.79\\(3.68$\times$)}
      & \cellcolor{redbg}\makecell{2.65\\(1.11$\times$)} \\
    524288 & 1 & 20 & 256 & fp16 & 1.0 &
      & 1.11 & 7.02
      & \cellcolor{greenbg}\makecell{0.99\\(0.89$\times$)}
      & \cellcolor{redbg}\makecell{10.5\\(1.50$\times$)}
      & \cellcolor{greenbg}\makecell{4.10\\(0.58$\times$)} \\
    524288 & 1 & 20 & 256 & fp16 & 1.15 &
      & 0.54 & 2.94
      & \cellcolor{redbg}\makecell{0.60\\(1.11$\times$)}
      & \cellcolor{deepredbg}\makecell{8.26\\(2.81$\times$)}
      & \cellcolor{lightgreenbg}\makecell{2.83\\(0.96$\times$)} \\
    \bottomrule
  \end{tabular}
  \centering
  \vspace{3pt}
  \caption{TBE training performance: MI350X forward and backward before/after optimization vs.\ B200 with representative production configs: B (batch size), T (number of tables), L (lookups per table), D (embedding dimension), Prec (precision), $\alpha$ (Zipf skew factor). Parenthetical values show latency relative to B200 (<1.0$\times$ means MI350X is faster).}
  \label{tab:tbe-results}
  \vspace{-10pt}
\end{table}

\subsubsection{Table Batched Embeddings (TBE)}

Table Batched Embeddings (TBE)~\cite{arxiv21-fbgemm} is a core primitive in Meta's recommendation workloads, responsible for looking up and aggregating embeddings from hundreds of tables with billions of rows.
TBE fuses multiple embedding lookups into a single kernel launch, amortizing kernel overhead and enabling memory access coalescing across tables.
The forward pass gathers embeddings based on sparse indices; the backward pass scatters gradients back to the embedding tables, requiring atomic accumulation for indices that appear multiple times.

\noindent \textbf{Backward optimization.}
We focus on the optimization of the custom backward kernel for MI350X's wavefront-64 architecture.
Key techniques include warp-per-row processing (each wavefront handles one embedding row), hand-tuned vectorized load/store for common embedding dimensions (64, 128, 192, 256), and wavefront-level accumulation that minimizes cross-lane communication.

\noindent \textbf{Results.}
Table~\ref{tab:tbe-results} summarizes TBE training performance across 12 production configurations spanning different batch sizes, table counts, sequence lengths, embedding dimensions, precisions, and index distributions.
The backward optimization delivers 1.87--3.32$\times$ speedup on MI350X, with larger gains on FP16 (2.5--3.3$\times$) than FP32 (1.9--2.8$\times$) due to better vectorization.
Compared to B200, MI350X with the optimized backward kernel is 1.35--1.96$\times$ faster on uniform distributions ($\alpha$=1.0), where memory bandwidth dominates.
On Zipf distributions ($\alpha$=1.15), where skewed access patterns cause cache thrashing, MI350X and B200 are comparable (0.83--1.16$\times$).

\subsection{Agentic Kernel Optimization Workflow}
\label{sec:agentic-tuning}

While GDPA and HSTU are widely shared across models, Meta's DLRM ecosystem contains a long tail of custom kernels tailored to individual models.
Manually tuning each kernel is labor-intensive and does not scale.
To address this, we developed an agentic AI workflow that encodes our tuning principles as executable skills, enabling automated kernel performance analysis and optimization.
This workflow builds upon Meta's KernelEvolve framework~\cite{isca26-kernelevolve}, which provides the underlying agent infrastructure and skill abstraction mechanisms.
While KernelEvolve focuses on the general systems architecture for agentic code optimization, our contribution lies in encoding AMD-specific tuning principles as production-ready skills and validating their effectiveness at scale on Meta's production DLRM training workloads.

The agentic workflow operates in three phases:
\begin{enumerate}
\item \textbf{Benchmark \& Profile:} Capture representative inputs from production workloads, reproduce the performance in a standalone benchmark, and leverage profiler tools (e.g., AMD rocprofv3~\cite{rocprofv3}, Triton Proton~\cite{triton-proton}) to identify the main bottlenecks such as low occupancy, poor cache utilization, or instruction stalls.
\item \textbf{Match \& Fix:} Map observed bottlenecks to a library of optimization skills, each encoding a tuning principle (e.g., PID swizzling, occupancy hints, atomic avoidance) with code transformation templates and applicability conditions.
\item \textbf{Generate \& Validate:} Produce candidate kernel modifications, benchmark them on representative inputs, and \rebuttal{verify numerical equivalence (within tolerance, automatically rejecting failures) and performance improvements on} a set of production workloads.
\end{enumerate}

\myparagraph{Case Study: Jagged Flash Attention.}
We illustrate the workflow with the jagged flash attention (JFA) kernel, which represents a challenging optimization target due to its irregular memory access patterns.
The JFA kernel, used in our ranking models, exhibited a $2\times$ slowdown on MI350X compared to B200 in its backward pass.
Profiling revealed that \texttt{tl.atomic\_add} operations dominate execution time, a known bottleneck on AMD's multi-XCD architecture (see Section~\ref{sec:cuda-kernels} for details).
The agent recognized this pattern, noting that the grouped query attention (GQA) mode causes multiple Q heads to race on shared \texttt{dK/dV} gradients.
Following the skill's guidance, the agent rewrote the kernel to accumulate \texttt{dK/dV} in registers per KV-head group and issue a single \texttt{tl.store} at the end, a tree-reduction pattern that eliminates atomic operations.
The optimized kernel achieved a $2.3\times$ speedup (32\,ms$\to$14\,ms), transforming MI350X from $2\times$ slower than B200 to 15--20\% faster, and delivering a 25\% improvement in end-to-end model training throughput.

Applying to dozens of production Triton kernels within Meta's DLRM ecosystem yields a success rate exceeding 80\%. Agent-generated optimizations achieve near-expert performance while reducing development time by an order of magnitude. For a comprehensive evaluation of the effectiveness across diverse kernel types, we refer readers to \cite{isca26-kernelevolve}.
By codifying expert knowledge into reusable skills with clear triggering conditions and transformation patterns, the workflow enables systematic exploration of the optimization space while ensuring that anti-patterns for AMD GPUs are automatically detected and addressed.

\rebuttal{\myparagraph{Generality.} Our workflow builds on Meta’s KernelEvolve~\cite{isca26-kernelevolve}, which encodes tuning expertise as reusable skills
across NVIDIA, AMD, and MTIA. The skill abstraction transfers across ISAs. The atomics-to-tree-reduction skill above has a direct analog on Meta's MTIA accelerators~\cite{artemis,athena}, whose PE array also handles contended atomics inefficiently: the same global-atomic reduction is rewritten as a per-PE local reduction followed by a global reduction over the on-chip reduction-engine network, yielding ${\sim}10\times$ (up to $65\times$) speedups on reduce-style kernels. Because the workflow needs only a portable tile-level kernel DSL and a compiler, which Google TPU (XLA/Pallas)~\cite{jouppi2017datacenter} and AWS Trainium (Neuron/NKI)~\cite{aws-trainium} also provide, these skills extend to a new accelerator as an incremental port rather than a from-scratch rebuild.}


\begin{figure*}[t]
  \centering
  \begin{subfigure}[t]{0.32\textwidth}
    \centering
    \includegraphics[width=\linewidth]{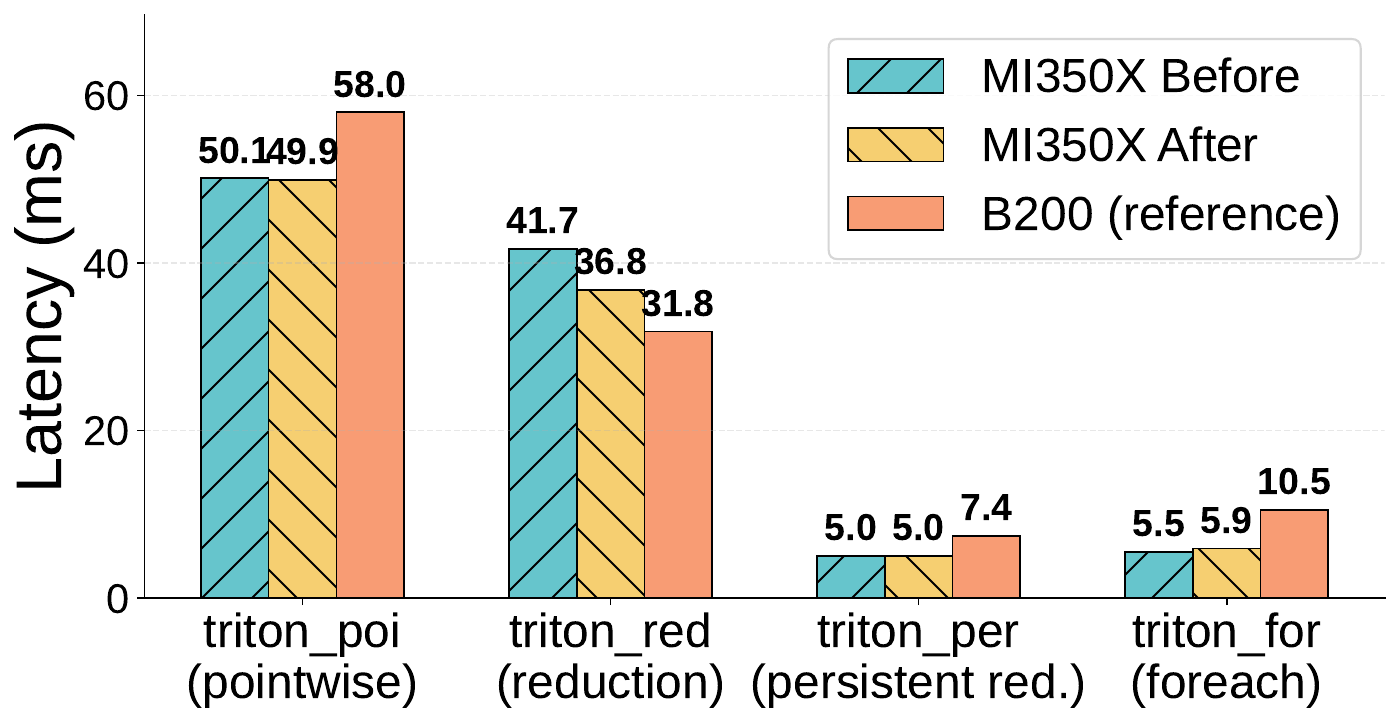}
    \caption{ResNet-50.}
    \label{fig:inductor-resnet50}
  \end{subfigure}\hfill
  \begin{subfigure}[t]{0.32\textwidth}
    \centering
    \includegraphics[width=\linewidth]{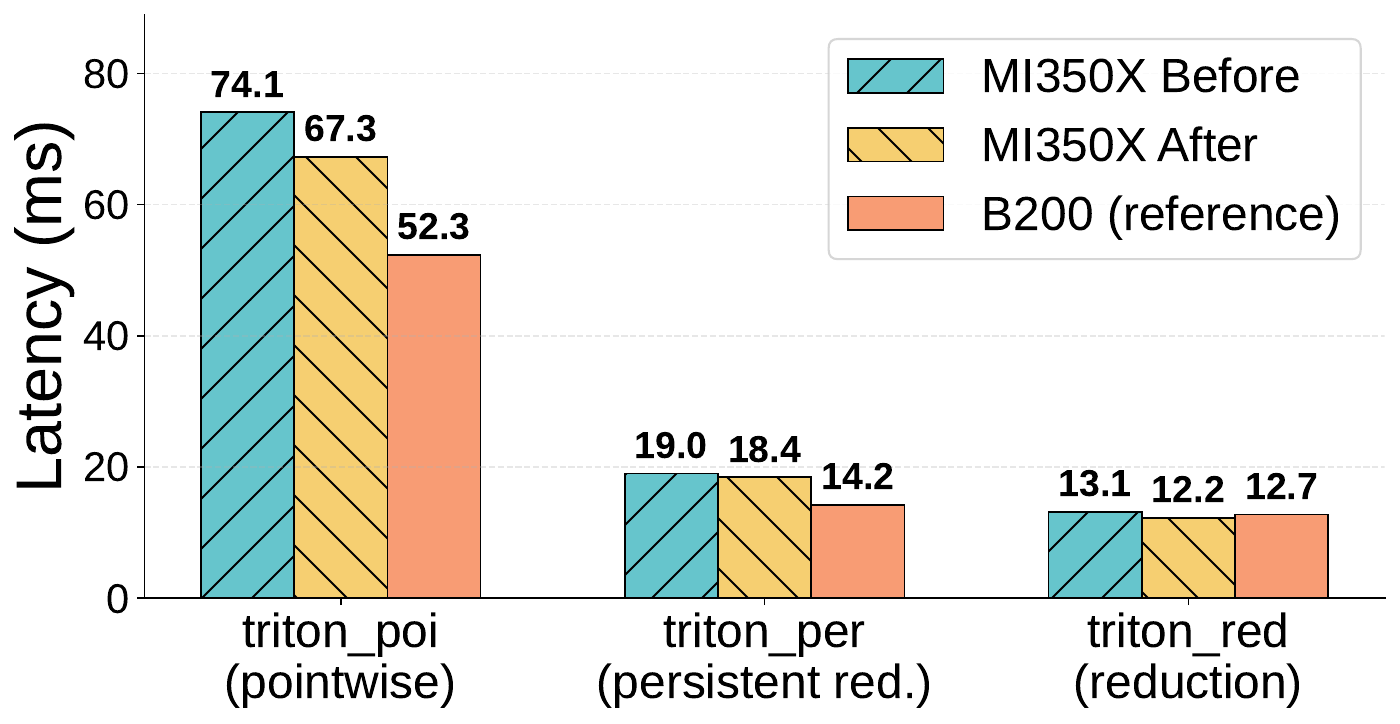}
    \caption{DenseNet-121.}
    \label{fig:inductor-densenet121}
  \end{subfigure}\hfill
  \begin{subfigure}[t]{0.32\textwidth}
    \centering
    \includegraphics[width=\linewidth]{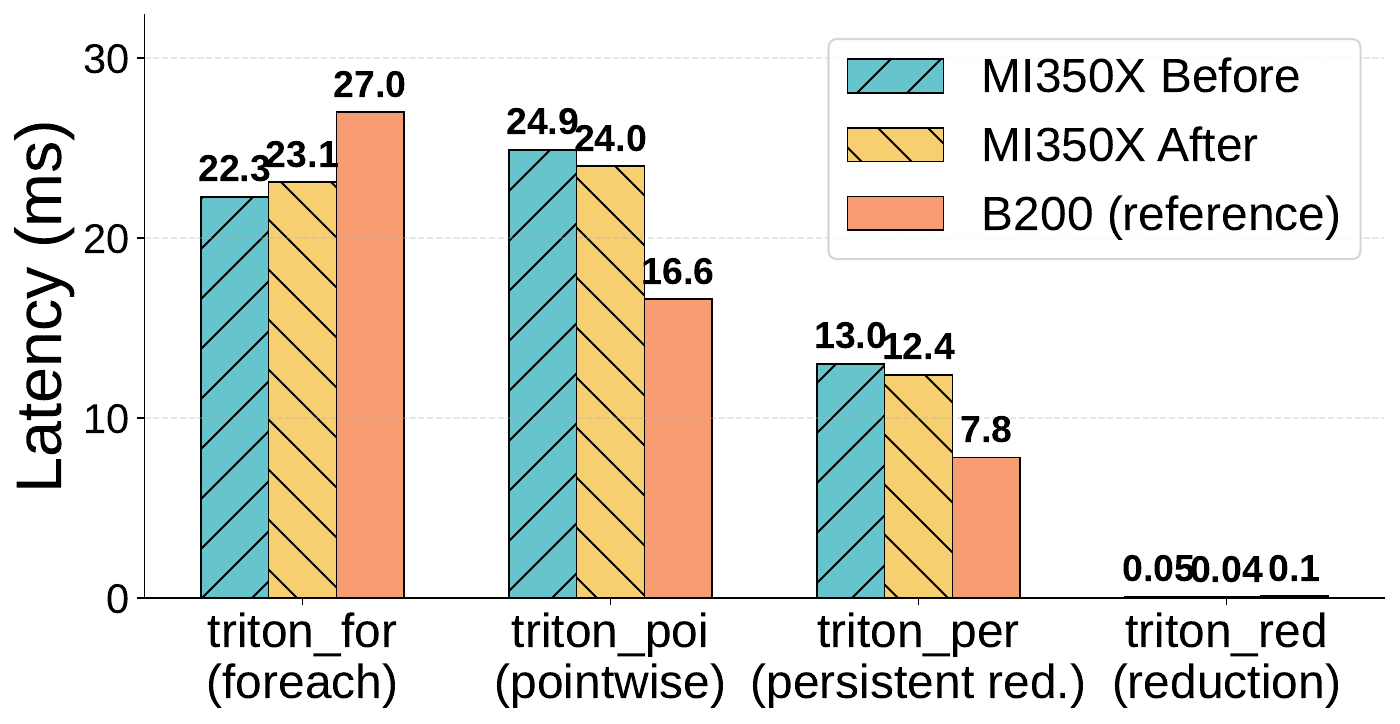}
    \caption{NanoGPT.}
    \label{fig:inductor-nanogpt}
  \end{subfigure}
  \caption{Triton Inductor kernel latencies by category for three representative models, showing MI350X before/after AMD optimizations compared to B200. Lower latency is better.}
  \label{fig:inductor-results}
  \vspace{-12pt}
\end{figure*}

\subsection{PyTorch Inductor Kernel Optimizations}
\label{sec:inductor-kernels}

PyTorch Inductor is the default backend for compiler \emph{kernel fusion}: transforming multiple eager operations into a single generated kernel launch to eliminate intermediate memory traffic, reducing bandwidth requirements by $2\times$--$5\times$ compared to unfused eager execution.

Inductor generates four main categories of kernels: \emph{pointwise} (\texttt{triton\_poi}) for elementwise operations (accounts for 40--60\% of the total kernel count), \emph{reduction} (\texttt{triton\_red}) for aggregations, \emph{persistent reduction} (\texttt{triton\_per}) for reductions fitting in registers or shared memory, and \emph{for-each} (\texttt{triton\_for}) for batched tensor operations.

Inductor's performance benefits are largely rooted in NVIDIA-specific optimizations refined over years of development and iterative tuning. Porting to alternative accelerators, such as AMD GPUs, necessitates more than mechanical adaptation; it requires re-evaluating core heuristics, expanding autotuning search spaces, and developing bespoke micro-architecture aware optimizations.

\subsubsection{Optimization Techniques}

We apply AMD-specific optimizations leveraging architectural insights from Section~\ref{sec:triton-kernels}:

    \noindent \textbf{Software pipelining.} We replace Python \texttt{range()} with Triton's \texttt{tl.range()} using \texttt{num\_stages=2}, enabling better memory-compute overlap. This reduces latency by 11.7\% for ResNet-50 reduction kernels.

    \noindent \textbf{Native Triton intrinsics} are used to replace library function calls where available: \texttt{tl.rsqrt()} instead of \texttt{libdevice.rsqrt()}, and \texttt{tl.minimum} / \texttt{tl.maximum} instead of helper functions.

    \noindent \textbf{Register configuration.} With 256 vector general purpose registers (VGPRs) per thread, we raise Inductor's register spill threshold from 16 to 32 for larger tiles and better instruction-level parallelism.

    \noindent \textbf{Expanded autotuning.}  We extend Inductor's autotuning configurations with AMD-specific tile sizes and occupancy hints.
    For example, we add non-square tiles ($64\times32$, $128\times16$, $128\times32$, $32\times512$) for 2D pointwise kernels.

\subsubsection{Model-Level Results}






We evaluate these optimizations on standard vision and language models.
Fig.~\ref{fig:inductor-results} shows per-category kernel latencies for three models trained on MI350X before/after optimization versus B200.
For \textbf{ResNet-50}~\cite{cvpr15-resnet}, reduction kernels improve 11.7\%, \rebuttal{reaching $1.16\times$ of B200 latency; other Triton inductor kernels outperform B200}.
For \textbf{DenseNet-121}~\cite{cvpr17-densent}, pointwise kernels improve 9.2\%; reduction improves 6.9\%, reaching B200 parity; persistent reduction improves 3.5\%. Overall kernel time is reduced by 4.2\%.
For \textbf{NanoGPT}~\cite{nanogpt}, persistent reduction improves 4.8\%; pointwise improves 3.8\%. Overall kernel time is reduced by 3.4\%.

In summary, AMD-specific optimizations deliver 6.9--11.7\% improvement for reduction, 3.5--4.8\% for persistent reduction, and up to 9.2\% for pointwise. While MI350X matches or exceeds B200 across several categories for CNNs, it is still 30\% slower on transformer Inductor kernels, indicating optimization headroom remains.

\section{Model-System Co-Design}
\label{sec:model}

AMD MI350X's large HBM capacity creates opportunities for model-system co-design, where training configurations and model architectures are jointly optimized to exploit the additional memory resources.
We explore two complementary directions: improving training throughput (examples/second) and improving model quality.

\subsection{Strategies for Training Efficiency}

\begin{figure}[t]
  \centering
  \begin{subfigure}[t]{0.235\textwidth}
    \centering
    \includegraphics[width=\linewidth]{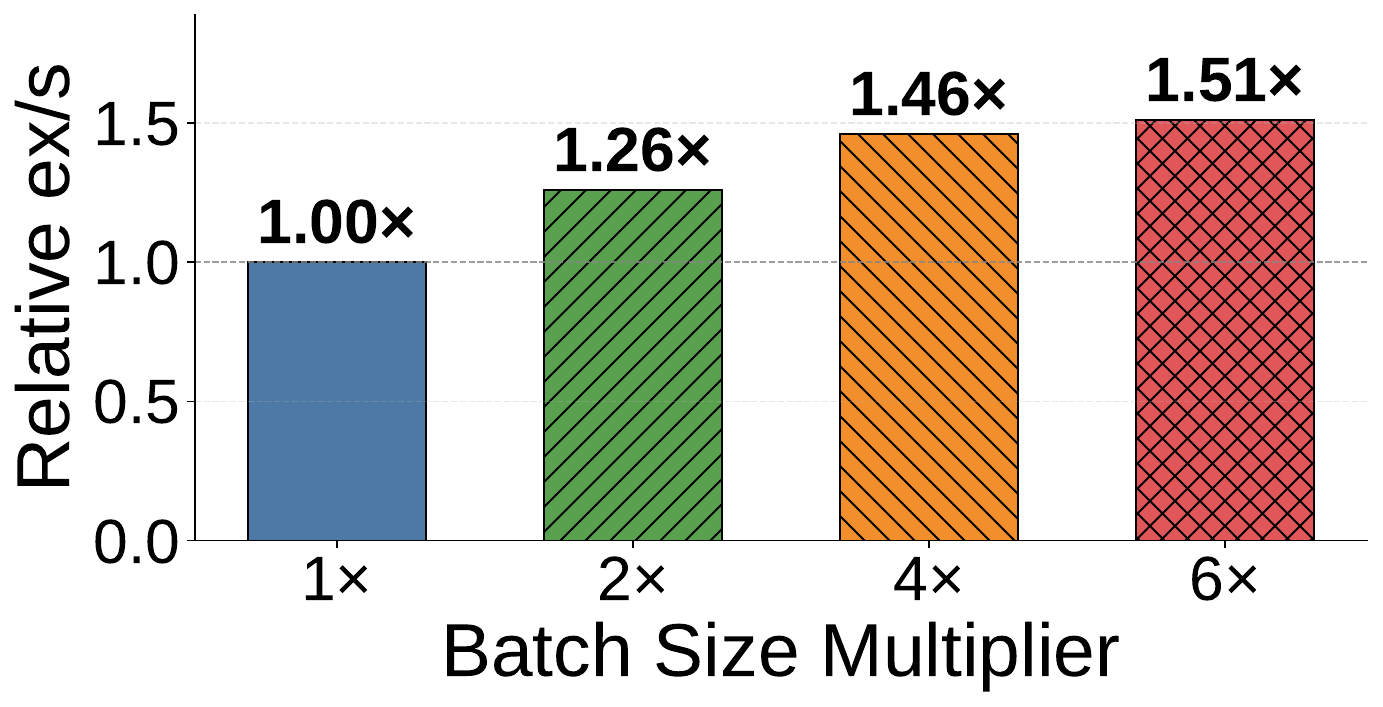}
    \caption{A model with increased batch sizes.}
    \label{fig:codesign-fb}
  \end{subfigure}\hfill
  \begin{subfigure}[t]{0.235\textwidth}
    \centering
    \includegraphics[width=\linewidth]{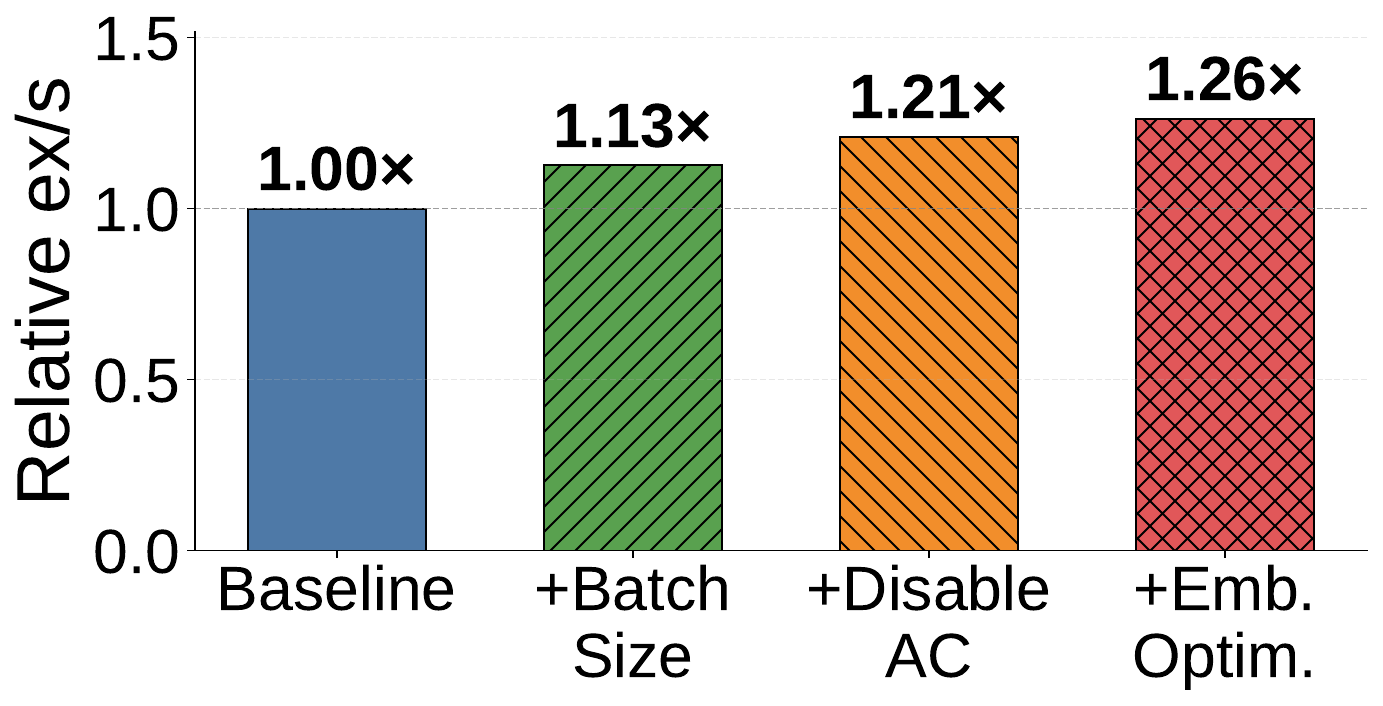}
    \caption{A model with co-design techniques.}
    \label{fig:codesign-ig}
  \end{subfigure}
  \caption{(a) Batch size scaling yields up to 51\% throughput improvement. (b) Combined co-design strategies (batch size increase, AC removal, embedding placement optimization) yield 26\% speedup.}
  \label{fig:model-codesign}
\end{figure}

\myparagraph{Increased batch size.}
Larger HBM capacity enables training with larger batch sizes, improving GPU utilization by amortizing kernel launch overhead and better saturating compute units.
For example, a friend recommendation model achieves 26\% speedup at $2\times$ batch size, 46\% at $4\times$, and 51\% at $6\times$ (Fig.~\ref{fig:codesign-fb}).
For a social ranking model migrated to MI350X, doubling the batch size yields 12--15\% throughput improvement over B200 baseline with the same batch size (B200 encounters out-of-memory errors at doubled batch size).

\begin{figure}[t]
\centering
\includegraphics[width=0.75\columnwidth]{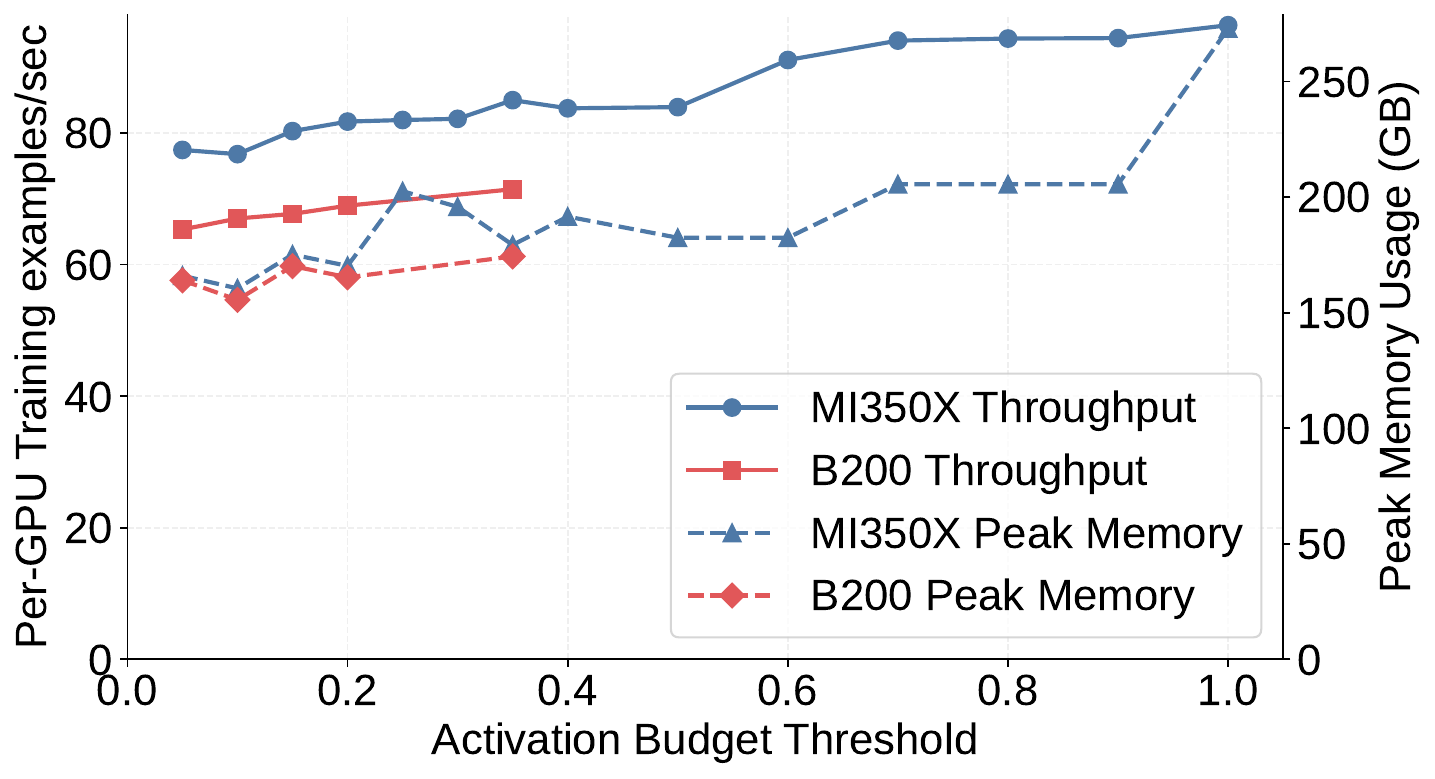}
\caption{Training throughput vs.\ HBM tradeoff on a GPT-2 1.3B model with activation checkpointing.}
\label{fig:gpt-ac}
\end{figure}

\myparagraph{Removal of activation checkpointing.}
Activation checkpointing (AC) trades memory for computation by recomputing activations during the backward pass instead of storing them.
With larger HBM, AC can be disabled for memory-constrained layers, eliminating redundant recomputation.
The activation memory budget controls how aggressively PyTorch checkpoints activations: a budget of 1.0 saves all activations (maximum memory, no recomputation), while lower values trigger selective recomputation to reduce memory at the cost of throughput.
Fig.~\ref{fig:gpt-ac} shows the tradeoff between activation memory budget and training throughput on a 1.3B-parameter GPT model. MI350X's larger HBM enables it to run without AC entirely, achieving 96.4~examples/second. In contrast, B200 must use aggressive AC (budget $\leq$ 0.35) just to fit, achieving 71.5~examples/second, a $1.35\times$ throughput disadvantage solely attributable to memory constraints.

\myparagraph{Dense parallelism strategy tuning.}
Additional HBM allows for relaxed sharding strategies, mitigating communication overhead.
For FSDP-based training~\cite{vldb23-fsdp}, moving from ZeRO-3 to ZeRO-2 or ZeRO-1~\cite{sc20-zero} reduces \texttt{AllGather} and \texttt{ReduceScatter} communication volume.
Switching to HSDP (Hybrid Sharded Data Parallelism) further hides communication latency by overlapping with computation.
For a large Ads foundation model at 1K GPU scale, HSDP provides 6\% speedup at the cost of 750\,MB additional memory per GPU, a tradeoff feasible only with HBM capacity headroom.

\myparagraph{Sparse embedding placement optimization.}
Recommendation models use large embedding tables that often reside in host memory due to GPU memory constraints.
With larger HBM, embedding tables can be moved from host to HBM, eliminating PCIe transfer bottlenecks.
Additionally, row-wise sharding can be replaced with table-wise or column-wise sharding, reducing \texttt{AlltoAll} communication and CPU/GPU synchronization overhead.
Experiments on a feed ranking model show $>$10\% speedup from embedding placement optimization, with the expectations for larger gains when combined with 2D embedding sharding strategies.

\myparagraph{Combined results.} Fig.~\ref{fig:codesign-ig} shows the combined effect of co-design strategies on a feed ranking model.
After porting to MI350X,
we increased the batch size, removed AC, and optimized embedding placement, resulting in a $1.26\times$ speedup.
In conjunction with kernel optimizations, the model realizes a total speedup of $1.5\times$ compared to the initial baseline.
These results demonstrate that combining multiple strategies delivers substantial efficiency improvements.

\subsection{Strategies for Model Quality}


\myparagraph{Model architecture upscaling.}
The MI350X’s expanded HBM capacity accommodates larger model architectures that would otherwise require additional GPU nodes. Increased memory footprint directly translates into superior model quality through several key avenues:
(1) \emph{Longer input sequences}: transformer-based recommendation models can leverage longer sequences to capture richer user context.
(2) \emph{Expanded embedding tables}: larger hash sizes reduce collision rates, thereby preserving higher-fidelity information within the learned representations.
(3) \emph{Increased dense complexity}: support for more dense layers and larger hidden dimensions enables the extraction of more expressive features.

\myparagraph{Results.}
For a feed ranking model, increasing the number of attention layers from 8 to 16 yields 0.08\% normalized entropy (NE)~\cite{workshop14-ne} improvement; further increasing to 20 layers yields 0.11\% NE improvement.\footnote{Achieving 0.1\% NE improvement in mature production models is non-trivial, gains of this magnitude typically require months of model iteration. At Meta's scale, such gains translate directly to meaningful improvements in user engagement.}
These architectural changes fit within AMD MI350X's memory capacity, whereas B200 would require cluster expansion.

\rebuttal{\myparagraph{Generality.} This co-design principle, matching the model to the accelerator's memory hierarchy, is not AMD-specific. It transfers directly to Meta's MTIA training accelerator (MTIA-300~\cite{athena}), whose larger HBM relative to the incumbent GPU is likewise exploited for larger batches and fewer trainers. The principle is direction-agnostic: on the SRAM-bound prior inference generation (MTIA-200~\cite{artemis}) it instead drives the model to \emph{shrink} its activation footprint to fit on chip, e.g., decomposing batched matmuls to remove redundant in-batch broadcasts cuts activation memory by ${\sim}25\%$, admitting $10$--$50\times$ larger embedding tables.}

\section{Evaluation}
\label{sec:eval}

The previous sections presented the efficacy of specific ML stack optimizations. This section assesses their collective impact on production recommendation models, illustrating how the cross-layer optimizations jointly maximize end-to-end system-level performance.

\subsection{Production Results}

Table~\ref{tab:eval-models} summarizes the job scale and measured throughput of ten representative production workloads from multiple product groups on MI350X and B200, using matched batch sizes.
GPU specifications follow the details in Section~\ref{sec:bg-amd}; readers should note the differences between our deployment and standard public offerings.
Evaluated models range from single-node (8 GPUs) to multi-node (128 GPUs), representing the diversity of production training scales.\footnote{Today, Meta's production DLRM workloads range from 8 GPUs (P25) to 64 GPUs (P75), with the P99 at 128 GPUs. Unlike LLMs trained as monolithic models at extreme scale, DLRM workloads comprise numerous distinct moderate-scale models serving different recommendation surfaces, collectively accounting for millions of job attempts.} \rebuttal{Here, \emph{Params} is the total parameter count (sparse embeddings plus dense layers), and \emph{Seq.\ len} is the longest input sequence a model ingests.}
The geometric mean performance ratio of 101.2\% demonstrates that MI350X \rebuttal{matches B200 throughput at equal GPU count and batch size} across the model suite.

\rebuttal{We use an equal number of B200 GPUs to give a straightforward comparison between same-generation hardware from both vendors (Section~\ref{sec:bg-amd}). As a widely adopted platform, Blackwell serves as a familiar anchor point for readers to evaluate our results.
The B200 baseline is not a strawman: as the long-established incumbent, it has received equal or greater optimization than the MI350X: years of model adoption, CUDA library tuning, and kernel development. The B200 keeps an edge on performance-per-watt, but power is a small share of capital-dominated total cost of ownership (TCO), so the MI350X's lower TCO drives provisioning.}


The production models also achieve higher per-GPU training throughput by leveraging larger HBM and batch sizes on MI350X, with gains in Model-D (98\% to 102\%), Model-E (94\% to 110\%), and Model-F (92\% to 95\%). For the kernel latency analysis in the next section, we intentionally use only the same batch sizes. While this underutilizes the MI350X, it provides consistent shapes needed to compare kernel-level performance between platforms.

\subsection{Kernel Category Analysis}

\begin{figure}[t]
  \centering
  \makebox[\linewidth][c]{%
    \includegraphics[width=1.05\linewidth]{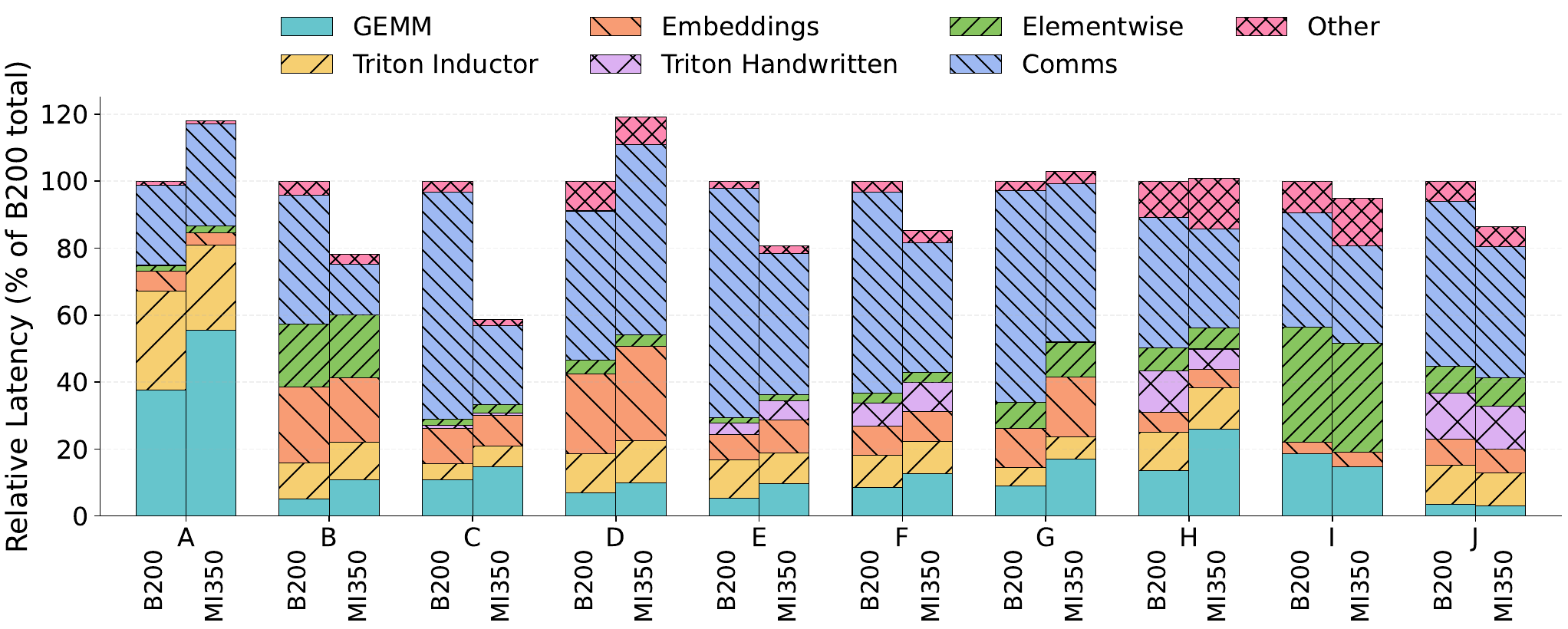}%
  }
  \caption{Kernel latency ratio (MI350X / B200) by category. For each model, the sum of kernel latencies on B200 is normalized to 100\%. This analysis does not consider multi-stream overlap.}
  \label{fig:trace-breakdown}
\end{figure}

To understand the sources of performance differences, we analyze GPU traces collected during training to categorize kernel latency by type: GEMM, Triton Inductor, CUDA embedding, elementwise, collectives, and hand-written Triton kernels.
Fig.~\ref{fig:trace-breakdown} shows the per-model breakdown of kernel latency ratio for each category.

\begin{table}[t]
  \centering
  \footnotesize
  \renewcommand{\arraystretch}{0.9}
  \setlength{\tabcolsep}{2pt}
  \begin{tabular}{@{}lrrrrrr@{}}
    \toprule
    Model & GPUs & \rebuttal{Seq.\ len} & \rebuttal{Params} & B200 ex/s & MI350X ex/s & MI350X/B200 \\
    \midrule
    A & 8 & \rebuttal{512} & \rebuttal{${\sim}100$M} & 921,714 & 754,936 & 82\% \\
    B & 8 & \rebuttal{256} & \rebuttal{${\sim}100$B} & 1,592,926 & 1,810,036 & 114\% \\
    C & 64 & \rebuttal{200} & \rebuttal{${\sim}200$B} & 728,001 & 884,216 & 121\% \\
    D & 64 & \rebuttal{1000} & \rebuttal{${\sim}300$B} & 774,324 & 755,221 & 98\% \\
    E & 64 & \rebuttal{200} & \rebuttal{${\sim}300$B} & 283,641 & 267,635 & 94\% \\
    F & 128 & \rebuttal{1000} & \rebuttal{${\sim}3.6$T} & 330,481 & 304,781 & 92\% \\
    G & 64 & \rebuttal{200} & \rebuttal{${\sim}30$B} & 544,537 & 499,066 & 92\% \\
    H & 64 & \rebuttal{24{,}600} & \rebuttal{${\sim}500$B} & 120,223 & 115,301 & 96\% \\
    I & 80 & \rebuttal{4096--8192} & \rebuttal{${\sim}1.9$T} & 546,072 & 558,127 & 102\% \\
    J & 24 & \rebuttal{700} & \rebuttal{${\sim}20$B} & 185,728 & 242,325 & 130\% \\
    \midrule
    \multicolumn{2}{@{}l}{\textbf{Geomean}} & \rebuttal{---} & \rebuttal{---} & --- & --- & \textbf{101.2\%} \\
    \bottomrule
  \end{tabular}
  \caption{End-to-end training throughput (examples/second) on MI350X vs B200 across ten production models. Higher is better.}
  \label{tab:eval-models}
  \vspace{-5pt}
\end{table}

\myparagraph{Where MI350X excels.}
\emph{Hand-written Triton kernels} show consistent MI350X advantages, delivering a 6\% to 52\% reduction in execution latency through the optimizations discussed in Section~\ref{sec:triton-kernels}.
Model-H benefits particularly from faster HSTU kernels, achieving $2.1\times$ speedup for the Triton category.
\emph{Embedding operations} show mixed results but generally favor MI350X ($0.61\times$--$1.04\times$), leveraging the CUDA kernel optimizations detailed in Section~\ref{sec:cuda-kernels}.

Overall, \emph{collective performance} is comparable between the two GPU types, with MI350X outperforming B200 in multi-host, small-message \texttt{AllToAll} collectives, specifically benefiting Models C and J.
Note that overlapped kernel execution presents a challenge for latency analysis: due to the multi-stream nature on GPUs, concurrent compute and communication kernels complicate the measurement of their individual contributions to end-to-end latency. For instance, collectives in Models E and F take longer to execute but are masked by compute kernels in the B200 traces, resulting in no net impact on throughput. Consequently, total kernel latency may overstate the true impact of collectives, leading to discrepancies between Fig.~\ref{fig:trace-breakdown} and measured throughput in Table~\ref{tab:eval-models}.

\myparagraph{Where B200 leads.}
\emph{GEMM operations} consistently favor B200, with MI350X showing $1.3\times$--$1.7\times$ higher latency.
MI350X lacks native hardware support for TF32, instead emulating them with three BF16 GEMMs at a performance penalty.
Model-A's overall 18\% performance deficit is largely attributable to TF32 GEMM operations.
While this is a notable limitation, modern models are increasingly moving toward BF16 or lower precision. We expect continued model modernization and improvements to emulated TF32 performance to further narrow the overall gap.
Additionally, cuBLAS outperforms hipBLAS due to its relative maturity, especially in the case of irregular matrix dimensions.
\emph{Triton Inductor kernels} show model-dependent results: some models achieve parity ($0.99\times$--$1.08\times$) while others show MI350X incurring $1.22\times$ higher latency than B200, reflecting ongoing efforts to refine Inductor and eliminate heuristics specifically tailored for NVIDIA GPUs (Section~\ref{sec:inductor-kernels}).

\subsection{\rebuttal{Operational Considerations}}
\rebuttal{Integrating a second GPU vendor into a previously NVIDIA-only fleet raised operational challenges in power, reliability, and observability. In addition to performance, operational maturity is just as critical to adopting a new vendor, and much of the tooling and fleet-management machinery we built generalizes to future accelerators.}

\rebuttal{\myparagraph{Reliability.} We track MI350X on the same metrics as our NVIDIA fleet (training-failure rate, unplanned unavailability, and hardware-swap rate), and it is steadily converging toward NVIDIA-fleet levels: the training-failure rate has more than halved from its peak as we resolved new at-scale failure modes, namely HBM RAS, multi-chiplet memory-access faults, and silent data corruption (SDC). SDC, the bit-flips that corrupt results without crashing, grows with cluster size and is the hardest to tame, so we address it end-to-end: a model-level checksum-and-replay check flags non-deterministic steps (distinguishing on-chip corruption from communication non-determinism); an all-pairs vote across a host's GPUs pinpoints the culprit device (the one that disagrees with all its peers) and auto-quarantines the host; and the same check runs during host qualification and burn-in to screen SDC-prone hosts before they enter the training pool, catching over $2\times$ the SDC of the vendor's field-health tool. These mitigations preserve correctness: across the tens of models we train on both vendors, results reach numerical parity (training-loss difference within $0.3\%$ and NE within ${\sim}10^{-4}$ at multi-thousand-GPU scale).}

\rebuttal{\myparagraph{Observability.} AMD telemetry initially lagged NVIDIA's mature stack, so we closed the gap: we integrated the efficiency metrics our NVIDIA fleet relies on (TFLOPS, SM/CU and memory utilization, throughput) and a hardware-agnostic trace-analysis interface (compute/communication breakdown, imbalance, overlap, and timeline analyses) that also enables side-by-side cross-vendor trace comparison (the basis for the apples-to-apples measurements here) and extends to MTIA.}

\rebuttal{\myparagraph{Power management.} Operating the MI350X against its excursion envelope rather than its steady-state limit is a fleet-level capability: it depends on platform provisioning beyond the OCP minimum and on tuning the control-loop response to production phase patterns. Section~\ref{sec:power-excursion} describes the mechanism and the resulting 3--6\% end-to-end gain.}

\section{Related Work}
\label{sec:related}

\myparagraph{Hardware Accelerators.}
Custom hardware accelerators exploit the structure of neural network computations. Meta's MTIA~\cite{mtia-gen1,meta2024mtia,mtia-blog} an in-house family of custom AI chips, designed to provide superior cost efficiency over commercial GPUs for the company's large-scale DLRM and GenAI workloads. Google's TPU~\cite{jouppi2017datacenter} employs systolic arrays for efficient matrix multiplication. Recent accelerators target emerging workloads: spiking neural networks~\cite{zhou2026spikex}, heterogeneous precision~\cite{zhou2024sysmol}, and processing-in-memory architectures that reduce the memory bandwidth bottleneck~\cite{kwon2019understanding}.

\myparagraph{Distributed Training Infrastructure.}
The LLM pretraining study on AMD MI300X GPUs~\cite{anthony2025amd} presents the transformer sharding strategy and the practical guidance for fault-tolerant distributed training.
Scaling DLRM training requires purpose-built infrastructure.
Zion~\cite{naumov2020zion} is a scale-up DLRM training platform that co-locates CPUs and accelerators to handle the memory capacity demands of embedding tables while leveraging GPU compute for dense operations. TorchRec~\cite{recsys22-torchrec} provides a PyTorch domain library with optimized distributed primitives, including automatic sharding planners that partition embedding tables across devices.

\myparagraph{Embedding Table Sharding and Placement.}
DLRMs present unique systems challenges due to their massive embedding tables, which can exceed hundreds of gigabytes and dominate both memory capacity and bandwidth requirements.
Intelligent partitioning of embedding tables across heterogeneous memory hierarchies is critical for training throughput.
RecShard~\cite{sethi2022recshard} formulates embedding placement as a mixed-integer linear program, leveraging per-feature access statistics—including categorical value frequency distributions and pooling factors—to achieve over 2.5$\times$ throughput improvement compared to size-based sharding. The key insight is that not all embedding rows are accessed equally; exploiting this skew enables fine-grained placement across GPU HBM, CPU DRAM, and SSDs. 2D sparse parallelism~\cite{arxiv25-2d-sparse} combines data and model parallelism for embeddings, enabling training to scale to thousands of GPUs.


\myparagraph{Kernel-Level Optimizations.}
Efficient GPU kernels are essential for embedding operations. FBGEMM~\cite{arxiv21-fbgemm} provides high-performance implementations of table batched embedding lookups, fused optimizers, and quantized operations optimized for recommendation workloads. These kernels exploit memory access coalescing, warp-level parallelism, and mixed-precision computation to maximize GPU utilization.
For inference, mixed-precision embeddings~\cite{li2024mpe} achieve 200$\times$ compression by learning per-feature precision assignments, grouping features by frequency to reduce the search space.
KernelEvolve~\cite{isca26-kernelevolve} provides an agentic framework for kernel optimization; we build specialized workflows that incorporates expert-tuning heuristics and architectural insights tailored for training workloads, simplifying cross-platform GPU kernel porting and optimization.


\section{Conclusion}
\label{sec:concl}

We present the first comprehensive study demonstrating that AMD MI350X GPUs achieve performance parity with NVIDIA B200 for production DLRM training through cross-stack optimizations spanning core ROCm libraries, kernel tuning, and model co-design. Our work addresses architectural differences between vendors---memory capacity, atomic performance, instruction sets---through principled algorithmic adaptations, hardware-aware autotuning, and HBM-conscious model co-design, providing a transferable framework for hardware-agnostic training at scale. Our optimizations have been upstreamed and open-sourced, establishing a foundation for multi-vendor ML infrastructure.


\balance
\bibliographystyle{IEEEtran}
\bibliography{ref}

\end{document}